\documentclass[reprint,showpacs,aps,amsmath,amssymb,prd,nofootinbib,nolongbibliography,floatfix]{revtex4-2} 
\usepackage{graphicx}
\usepackage{dcolumn}
\usepackage{bm}
\usepackage{ulem} 
\usepackage{float}
\usepackage[usenames]{color}

\usepackage{times}
\usepackage{mathptmx}

\usepackage{amssymb,amsmath}
\usepackage{latexsym}

\definecolor{linkcolor}{rgb}{0.0,0.3,0.5}
\usepackage[unicode, colorlinks=true, linkcolor=linkcolor, citecolor=linkcolor, filecolor=linkcolor, urlcolor=linkcolor, linktocpage, breaklinks]{hyperref}
\usepackage[all]{hypcap}
\usepackage[T1]{fontenc}
\usepackage[utf8]{inputenc}
\usepackage[usenames,dvipsnames]{xcolor}
\hypersetup{colorlinks=true,citecolor=romared,linkcolor=romared,urlcolor=romared}

\definecolor{romared}{RGB}{142,0,28}

\usepackage{cancel}
\newcommand{\beq}{\begin{equation}}
\newcommand{\eeq}{\end{equation}}
\newcommand{\beqn}{\begin{eqnarray}}
\newcommand{\eeqn}{\end{eqnarray}}
\newcommand{\pa}{\partial}

\newcommand{\varep}{\varepsilon}

\def\zeroD{\stackrel{(0)}{\mathstrut D}\hspace{-1.4mm}}
\def\zeroDelta{\stackrel{(0)}{\mathstrut \Delta}}

\def\2pi{2\pi}

\begin{document}
\title{Properties of scalar wave emission in a scalar-tensor theory with kinetic screening}
\author{Masaru Shibata$^{1,2}$}
\email{masaru.shibata@aei.mpg.de}
\author{Dina Traykova$^{1}$}
\email{dina.traykova@aei.mpg.de}
\affiliation{$^1$Max Planck Institute for
  Gravitational Physics (Albert Einstein Institute), Am M{\"u}hlenberg 1,
  Potsdam-Golm 14476, Germany}
\affiliation{$^2$Center for Gravitational Physics and Quantum-Information,
  Yukawa Institute for Theoretical
  Physics, Kyoto University, Kyoto, 606-8502, Japan}

\begin{abstract}
We study numerically the scalar wave emission by a non-spherical oscillation of neutron stars in a scalar-tensor theory of gravity with kinetic screening, considering both the monopole and quadrupole mode emission.
In agreement with previous results in the literature, we find that the monopole is always suppressed by the screening effect, regardless of the size of the screening radius, $r_{\rm sc}$.
For the quadrupole mode, however, our analysis shows that the suppression only occurs for screening radius larger than the wavelength of scalar waves, $\lambda_{\rm wave}$, but not for $r_{\rm sc} < \lambda_{\rm wave}$.
This demonstrates that to fully understand the nature of this theory, it is necessary to study other more complex systems, such as neutron star binaries, considering a wide range of $r_{\rm sc}$ values.
\end{abstract}

\maketitle

\section{Introduction}

The ample evidence for the current accelerated expansion of the Universe has hinted at the existence of some new physics at cosmological scales \cite{Riess:1998cb,Perlmutter:1998np,komatsu2009five,Komatsu_2011,Abbott:2018wog,Hinshaw:2012aka,Aghanim:2018eyx,Alam:2016hwk}.
One of the simplest modifications to general relativity (GR), which can provide a possible explanation of this phenomenon, is the so-called scalar-tensor theories, where an additional scalar degree of freedom is minimally (e.g. quintessence~\cite{Ratra:1987rm,Wetterich:1987fm,Ferreira:1997hj,Caldwell:1997ii}; see also Refs.~\cite{Copeland:2006wr,Tsujikawa:2013fta} for reviews) or non-minimally coupled to the gravitational metric (see Refs.~\cite{Bergmann:1968ve, fujii_maeda_2003,amendola_tsujikawa_2010, Clifton:2011jh} for a review on scalar-tensor gravity).
On cosmological scales, it is possible to measure and constrain physical parameters that capture this novel behaviour \cite{Abell:2009aa,Font-Ribera:2013rwa,Spergel:2013tha,Aghamousa:2016zmz,Hounsell:2017ejq}, showing that modifications to GR that can account for the observed accelerated expansion of the Universe 
on these scales with the dark sector whose  density is of the order of the critical density, $\rho_c$.
This means that we can expect similar deviations on small scales too. However, Solar System \cite{Will:1993tegp,Will:2005va} and binary pulsar \cite{Damour:1992mar,Kramer:2006nb,Freire:2012mg,Kramer:2021jcw, Zhao:2022vig} tests show no violations of the predictions of GR there.
In addition, radio observations of pulsars (neutron stars) accompanying white dwarfs constrain the emissivity of scalar-type gravitational waves (hereafter referred to simply as scalar waves), and thus, the parameter space for some scalar-tensor theories has been significantly limited~\cite{Will:1989bin,Freire:2012rpw,Zhao:2022vig}.
More recently, consistency with GR has also been shown by null tests with gravitational-wave observations \cite{LIGOScientific:2016lio,LIGOScientific:2018dkp,LIGOScientific:2019fpa,LIGOScientific:2020tif,LIGOScientific:2021sio}.

One possible solution to this problem is employing an appropriate screening mechanism, by which the effects of the scalar field are suppressed on local scales so that GR phenomena can be reproduced, while on cosmological scales, modifications to GR remain appreciable. Some well-studied examples of this behaviour are the chameleon \cite{Khoury:2003aq}, symmetron \cite{Hinterbichler:2010es}, and Vainshtein \cite{Vainshtein:1972sx,Deffayet:2001uk,Nicolis:2008in} screening (see also  Refs.~\cite{Brax:2013ida, Joyce:2014kja, Koyama:2015vza} for reviews). 
Even though screening effects have been studied extensively in a range of simplified scenarios, such as weak-gravity and spherical symmetry approximations (see, e.g., Refs.~\cite{Babichev:2009ee,Babichev:2010jd,deRham:2012fw,Babichev:2013pfa,Bloomfield:2014zfa,Crisostomi:2017lbg,Nakamura:2018gxf,Dar:2018dra,Burrage:2018xta,Burrage:2020bxp,Burrage:2021nys}), they are not so well-understood in strongly self-gravitating and dynamical environments, such as the dynamical neutron star spacetime.
For example, the emission mechanism of scalar waves has not been yet well-understood. 
In order to fully characterise the screening effect in dynamical spacetimes, for which no linearization or symmetry of the system can be employed, numerical relativity (NR), by which the solution of the fully non-linear systems can be obtained, is needed.

NR simulations of compact objects in scalar-tensor theories with a kinetic screening effect have been performed in a few recent studies~\cite{Bernard:2019fjb,Bezares:2021yek,Bezares:2021dma,Dima:2021pwx,Lara:2021piy,Lara:2022gof}, some of which report a non-trivial nature of the scalar-wave emission. 
In particular, in Ref.~\cite{Bezares:2021dma}, the authors find that the quadrupole scalar wave emission may not be screened in the case of a binary neutron star inspiral. 
This study focuses on the cases with a small screening radius ($\alt 140$\,km), which is smaller than the wavelength of gravitational and scalar waves. 
We argue here that, in a such setting, the screening effect may not be significant and one could expect different behaviour when larger screening radii, which are more realistic, are considered.

In this paper, we study numerically the emission of scalar waves from non-spherically oscillating neutron stars in the same scalar-tensor theory employed in Ref.~\cite{Bezares:2021dma}. 
It has been shown in Ref.~\cite{Nakao:2000ug} that scalar waves in a scalar-tensor theory of gravity can be detected by interferometers in the same way as gravitational waves. Their analysis, done in the framework of the Brans-Dicke theory shows that, for a simple Michelson interferometer, the antenna sensitivity pattern depends strongly on the frequency of the scalar gravitational waves, with essentially the same features as those of the tesnosr mode of GWs. Thus showing that as long as the dependence of the antenna sensitivity pattern on the wave length of scalar waves is taken into account in the same way as for the tensor modes, scalar waves would be detectable in the case of a scalar-tensor theory. Therefore in this work we treat both scalar and tensor modes as gravitational waves.

Our NR simulation is performed in the Jordan frame in contrast to previous works~\cite{Bezares:2021dma,Bezares:2021yek}, which employ the Einstein frame instead.
Doing this has three advantages:
(i) the equations for hydrodynamics are not changed and have a conservative form, same as in GR; 
(ii) the gravitational and scalar waves are extracted independently from the spacetime metric and the scalar field, respectively;
and (iii) unlike the Einstein frame case, the Jordan frame metric couples universally to the matter fields and so observables can also be computed in the same way as one does in GR.
We will show that if the screening radius is larger than the wavelength of scalar waves, the screening effect on the scalar waves (i.e., the suppression of the scalar wave emission) is always significant irrespective of the multipoles considered. 

The paper is organised as follows. 
In Sec.~\ref{sec:equations} we summarise the basic equations that we employ. 
Section~\ref{sec:initial_cond} presents a formulation for computing equilibrium and quasi-equilibrium states, necessary for the initial conditions in NR simulation. 
Section~\ref{sec:spher_NS} presents numerical solutions of $1.4M_\odot$ spherical neutron stars and summarises the properties of a neutron star spacetime in the presence of the kinetic screening.
In Sec.~\ref{sec:non-spher_NS} we explore the non-spherical oscillation of neutron stars obtained in Sec.~\ref{sec:spher_NS}, in particular focusing on the generation and propagation of quadrupole scalar waves.
Finally, we discuss our results and summarise our conclusions in Sec.~\ref{sec:discussion}.

Throughout this paper, we use the units of $c=1=\hbar$, where $c$ and $\hbar$ denote the speed of light and the reduced Planck constant, respectively.
In these units the Planck length, $\ell_{\rm p}:=G^{1/2}=1.616 \times 10^{-33}$\,cm and the Planck mass, $M_{\rm p}:=G^{-1/2}=2.176\times 10^{-5}$\,g.  
The subscripts $a$ and $b$ denote the spacetime tensor components, and $i$, $j$, and $k$ denote the spatial components.

\section{Basic equations}
\label{sec:equations}

In this work we consider a scalar-tensor theory with kinetic screening, in which the action in the so-called Jordan frame is
given by~\cite{Jordan:1949zz,Fierz:1956zz,Brans-Dicke,Will:1993tegp},
\beqn
S &=& \frac{1}{16 \pi G} \int d^4x \sqrt{-g} \phi\,
\left[{\cal R} +\left(\frac{3}{2} + \frac{\hat K}{\alpha_{\rm s}^2}\right)
 g^{ab}  \frac{\nabla_a \phi\, \nabla_b \phi}{\phi^2} \right] \nonumber \\
&&+S_{\rm matter}(\chi_{\rm matter},g_{ab})\,.
\label{eqn:action}
\eeqn
The corresponding action in the Einstein frame can be found in, e.g., Refs.~\cite{Chiba:1999ka,Bezares:2021dma}.
Here ${\cal R}$ and $\nabla_a$ are the Ricci scalar and covariant derivative associated with the spacetime metric $g_{ab}$, $\phi (>0)$ is the gravitational scalar field and $\hat K$ is a function of the canonical kinetic term of the scalar field, $X$. 
$S_{\rm matter}$ is the action of the perfect fluid, with $\chi_{\rm matter}$ representing 
the matter fields.
The kinetic term of the scalar field is defined as,
\beqn
X=\bar g^{ab} \bar\nabla_a \bar\varphi \bar\nabla_b\bar\varphi
=\phi^{-1} g^{ab} \nabla_a \bar\varphi \nabla_b\bar\varphi\,,
\eeqn
where $\bar g_{ab}$ is the spacetime metric in the Einstein frame, 
$\bar\nabla_a$ is its covariant derivative, 
$\bar\varphi=\ln\phi/\sqrt{16\pi G\alpha_{\rm s}^2}$, and $\alpha_{\rm s}$ is
a coupling constant. 
Following Ref.~\cite{Bezares:2021dma}, we consider the case,
\beq
\hat K(X)=-\frac{1}{2} + \frac{\gamma_1}{4\Lambda^4} X - \frac{\gamma_2}{8 \Lambda^8} X^2 \cdots,
\eeq
where $\Lambda$ is the strong-coupling scale (i.e., $\lambda:=\Lambda^{-1}$ determines the length scale of screening), and $\gamma_1$ and $\gamma_2$ are constants of order unity. 
Here we choose $\gamma_1=0$ and $\gamma_2=1$ 
as it has been shown (see Refs.~\cite{Bezares:2020wkn,terHaar:2020xxb}) that this is a necessary condition for having a well-posed initial value formulation, as well as screening static solutions.
Screening is expected to occur in the strong field zone, where $X > \Lambda^{4}$ is satisfied.  
We suppose that $\phi \rightarrow 1$ (i.e., $\bar\varphi\rightarrow 0$ and $X \rightarrow 0$) for $r \rightarrow \infty$.

For $\gamma_1=\gamma_2=0$ this theory is equivalent to the Fierz-Jordan-Brans-Dicke (FJBD) theory~\cite{Jordan:1949zz,Fierz:1956zz,Brans-Dicke}, with Brans-Dicke parameter of the form,
\beq
\omega(X):=-\frac{3}{2} - \frac{\hat K(X)}{\alpha_{\rm s}^2}\,,
\label{omega}
\eeq
with $X=0$, so that $\omega(X)=-\frac{1}{2}\left(3-{\alpha_{\rm s}}^{-2}\right)$.

Then the basic equations for the geometry, scalar field, energy momentum tensor, $T_{ab}$, and rest-mass continuity are as follows,
\beqn
&&G_{ab}= 8\pi G \phi^{-1} T_{ab} \nonumber \\
&&~~~~~~-\left(\frac{3}{2}+\frac{\hat K}{\alpha_{\rm s}^2}\right)\phi^{-2}
\biggl[(\nabla_a\phi) \nabla_b \phi-\frac{1}{2}g_{ab}
  (\nabla_c\phi)\nabla^c\phi \biggr] \nonumber \\
&&~~~~~~
-\frac{X}{ \alpha_{\rm s}^2\phi^2} \frac{\pa \hat K}{\pa X}\nabla_a \phi \nabla_b \phi
+\phi^{-1} (\nabla_a\nabla_b \phi - g_{ab} \Box_g \phi),\label{eq:JBD1} \\
&&\nabla^a \left( F \nabla_a \phi\right)
= 8\pi G \alpha_{\rm s}^2 T , \label{eq:JBD2} \\
&&\nabla_a T^a_{~b}=0,\label{eq:JBD3}\\
&&\nabla_a (\rho u^a)=0, 
\eeqn
where $G_{ab}$ is the Einstein tensor associated with $g_{ab}$, $T=T_a^{~a}$, $u^a$ is the fluid four velocity, $\rho$ is the rest-mass density, and
\beqn
F:=-2 \frac{\pa (X \hat K)}{\pa X} 
=1-\gamma_1 \frac{X}{\Lambda^4} + \frac{3\gamma_2}{4}
\frac{X^2}{\Lambda^8} +\cdots. 
\eeqn
To derive Eq.~\eqref{eq:JBD2}, we used the trace of Eq.~\eqref{eq:JBD1},
\beqn
-{\cal R}&=&8\pi G \phi^{-1} T +
\left( \frac{3}{2} + \frac{\hat K}{\alpha_{\rm s}^2}
-\frac{X}{\alpha_{\rm s}^2} \frac{\pa \hat K}{\pa X}\right)
\frac{(\nabla_a \phi) \nabla^a \phi}{\phi^2} \nonumber \\
&& -\frac{3}{\phi} \Box_g \phi\,, 
\eeqn
where $\Box_g=\nabla_a \nabla^a$.

For $T_{ab}$, we consider the stress-energy tensor for a perfect fluid,
\beq
T_{ab}=(\rho + \rho \varep + P)u_a u_b + P g_{ab}, 
\eeq
where $\varep$ and $P$ are the specific internal energy and pressure of the fluid. 
In the Jordan frame the fluid matter is coupled only to the gravitational field, as seen in Eq.~\eqref{eq:JBD3}.
Hence, the equations for the perfect fluid are the same as those in GR in this frame.

The basic equations in the 3+1 formulation for the gravitational field are derived simply by contracting $n^a n^b$, $n^a \gamma^b_{~i}$, and $\gamma^a_{~i} \gamma^b_{~j}$ with Eq.~\eqref{eq:JBD1}.
Here, $\gamma_{ab}=g_{ab}+n_a n_b$ denotes the spatial metric, and $n^a$ is the unit normal to the spatial hypersurfaces.  The 3+1 form of the scalar field equation is derived from Eq.~\eqref{eq:JBD2} by defining $\Pi:=-n^a \nabla_a \phi$ or $\hat\Pi:=-F n^a \nabla_a \phi$.

The evolution of the scalar field and its conjugate momentum have the following form,
\beqn
(\pa_t -\beta^k \pa_k)\phi &=& -\alpha \Pi,\label{eq:JBD5a} \\
(\pa_t -\beta^k \pa_k)\hat \Pi &=& -D_i\left(\alpha F D^i \phi \right) 
+\alpha K \hat\Pi \nonumber \\
&&+8\pi G \alpha \alpha_{\rm s}^2 T\,,
\label{eqn:field_evol}
\eeqn
where $D_i$ is the covariant derivative with respect to $\gamma_{ij}$.
In terms of $\Pi$ and $\phi$, $X$ can be written as,
\beq
X=\frac{1}{16\pi G \alpha_{\rm s}^2 \phi^3}\left[
(D_k\phi) D^k\phi - \Pi^2\right]\,. \label{eqX}
\eeq
From these one can also obtain an algebraic equation for $X$,
\beq
f(X):=X-\frac{1}{16\pi G \alpha_{\rm s}^2 \phi^3}
\left[(D_k\phi) D^k\phi - \frac{\hat\Pi^2}{F(X)^2}\right]=0\,. \label{eqn:X2}
\eeq
For a detailed description of the 3+1 equations of the system, we refer the reader to Appendix~\ref{app:3+1}.

The evolution equations for the gravitational fields are solved numerically in the Baumgarte-Shapiro-Shibata-Nakamura (BSSN) formalism~\cite{Shibata:1995we,Baumgarte:1998te} with the moving-puncture gauge~\cite{Campanelli:2005dd,Baker:2005vv}, as done in
Ref.~\cite{Shibata:2013pra}.
In particular, we evolve the conformal factor $W:= \psi^{-2}$ (with $\psi:=({\rm det} \gamma_{ij})^{1/12}$), the conformal metric
$\tilde{\gamma}_{ij} := \psi^{-4} \gamma_{ij}$, the trace part of the extrinsic curvature $K$, the conformally weighted trace-free part of the extrinsic curvature $\tilde{A}_{ij}:=\psi^{-4}(K_{ij}-K\gamma_{ij}/3)$ (with $K_{ij}$ -- the extrinsic curvature), and the auxiliary variable $\tilde{\Gamma}^i := - \partial_j \tilde{\gamma}^{ij}$.
Introducing the auxiliary variable $B^i$ and a parameter $\eta_s$, which is typically set to be $\sim M^{-1}$, $M$ being the total mass of the system\footnote{We note that the total mass includes a contribution both from the ADM mass and the scalar charge (see the tensor mass in Sec.~\ref{sec:initial_cond}).}, we employ the moving-puncture gauge in the form of~\cite{Bruegmann:2006ulg},
\begin{eqnarray}
 ( \partial_t - \beta^j \partial_j ) \alpha &=& - 2 \alpha K , \\
 ( \partial_t - \beta^j \partial_j ) \beta^i &=& (3/4) B^i , \\
 ( \partial_t - \beta^j \partial_j ) B^i &=& ( \partial_t - \beta^j
  \partial_j ) \tilde{\Gamma}^i - \eta_s B^i ,
\end{eqnarray}
where $\alpha$ and $\beta^i$ are the lapse function and shift vector, respectively. 

The spatial derivative is evaluated by a fourth-order central finite difference scheme, except for the advection terms, which are evaluated by a fourth-order non-centred finite difference. 
For the time evolution, we employ a fourth-order Runge-Kutta method (see Ref.~\cite{Yamamoto:2008js}). 
We use the same scheme for the evolution of the scalar field as for the tensor, because the structure of the equations is essentially the same.

To solve the hydrodynamics equations, we evolve $\rho_* := \rho \alpha u^t W^{-3}$, $\hat{u}_i := h u_i$, and $e_* := h \alpha u^t - P / (\rho \alpha u^t )$, with $h$ being the specific enthalpy. 
The advection terms are handled with a high-resolution shock capturing scheme of a third-order piecewise parabolic interpolation for the cell reconstruction. 
For the equation of state (EOS), we decompose the pressure and the specific internal energy into cold and thermal parts as,
\begin{equation}
 P = P_{\rm cold} + P_{\rm th} \; , \; \varepsilon = \varepsilon_{\rm cold} + \varepsilon_{\rm th}\,.
\end{equation}
Here, $P_{\rm cold}$ and $\varep_{\rm cold}$ are functions of $\rho$, and their forms are determined by nuclear-theory-based
zero-temperature EOSs.
Specifically, the cold part of both variables are determined using the piecewise polytropic version (see, e.g.,
Ref.~\cite{Hotokezaka:2013mass}) of the APR4 EOS~\cite{Akmal:1998cf}, for which the maximum mass of the neutron stars in GR is $\approx 2.2M_\odot$. 

Then the thermal part of the specific internal energy is defined from $\varepsilon$ as $\varepsilon_{\rm th} := \varepsilon - \varepsilon_{\rm cold}$.
Because $\varepsilon_{\rm th}$ vanishes in the absence of shock heating, $\varepsilon_{\rm th}$ is regarded as the finite-temperature part (and thus, this part is minor in the present study). 
The thermal pressure is determined by a $\Gamma$-law EOS,
\beq
 P_{\rm th} = ( \Gamma_{\rm th} - 1 ) \rho \varepsilon_{\rm th}\,,
\eeq
and we choose $\Gamma_{\rm th}$ equal to 1.8, following Refs.~\cite{Shibata:2013pra,Hotokezaka:2013mass}. 


\section{Formulation for initial conditions}
\label{sec:initial_cond}
Here we outline the formulation for computing quasi-equilibrium configurations for a binary in a circular orbit with angular velocity $\Omega$ following Refs.~\cite{Isenberg:2008wat,Wilson:1995icn,Wilson:1996ty}.
This description is also valid for computing static spherical stars with $\Omega=0$.

To derive quasi-equilibrium configurations, for simplicity, we assume the conformal flatness of the three metric, such that
\beqn
\gamma_{ij}=\psi^4 f_{ij}\,,
\label{eq:conflat}
\eeqn
where $f_{ij}$ is the flat spatial metric, and employ the conformal thin-sandwich prescription.
We also impose the maximal slicing $K=0$.
For integrating the hydrodynamics equations, we assume the presence of a helical Killing vector, $\xi^a=(\pa_t + \Omega \pa_\varphi)^a$.
For the fluid part, the basic equations in the Jordan frame are the same as those in GR.
Thus, assuming that the velocity field is irrotational, the first integral of the hydrodynamics equations is readily determined in the same manner as those in GR~\cite{Teukolsky:1998sh,Shibata:1998um}.

The basic equations for the tensor field are obtained from the Hamiltonian and momentum constraints, together with the evolution equation for $K$ (see Appendix~\ref{app:3+1}) under the maximal slicing condition, $K=0=\pa_t K$. 
Except for the modifications introduced by the presence of the scalar field, $\phi$, the equations are again the same as in GR.
The Hamiltonian and momentum constraints are written as,
\beqn
\zeroDelta \hspace{-1mm} \psi &=& -2\pi G \phi^{-1} \rho_{\rm h} \psi^5 
-\frac{1}{8} \tilde A_{ij} \tilde A^{ij} \psi^{5}
 \nonumber \\
 &-&\frac{\psi^5}{8}\left[\frac{\omega}{\phi^{2}}
   \left\{\Pi^2 + (D_i\phi)D^i\phi\right\}
  +2\phi^{-1} D_i D^i \phi
  \right. \nonumber \\
 && \left.
  \hskip 2.cm
-\frac{2 \Pi^2}{\alpha_{\rm s}^2 \phi^2}X \frac{\pa \hat K}{\pa X}
 \right]\,,
\label{eq:Ham1}
\eeqn
and
\beqn
\zeroD_i (\psi^6\tilde A^{i}_{~j})&=&\psi^6\left[
8\pi G\phi^{-1} J_j + \left(\omega 
-\frac{X}{\alpha_{\rm s}^2}\frac{\pa \hat K}{\pa X}
\right)\phi^{-2} \Pi \zeroD_j\phi
\right. \nonumber \\
&&\left.~~~+\phi^{-1}(\zeroD_j \Pi - \tilde A^i_{~j} \zeroD_i \phi)\right],
\label{eq:Mom1}
\eeqn
respectively. 
Here $\zeroDelta$ and $\zeroD_i$ are the Laplacian and covariant derivatives with respect to $f_{ij}$, $\rho_{\rm h}:=T_{ab}n^a n^b$, and $J_i:=-T_{ab}n^a \gamma^b_{~i}$. 
$\tilde A_{ij}$ is the trace-free conformal extrinsic curvature, satisfying $K_i^{~j}=\tilde A_i^{~j}$ for $K=0$.
The equation for $\tilde A_{ij}$ can be obtained from the evolution equation for $\gamma_{ij}$ with Eq.~\eqref{eq:conflat} and has the form,
\beqn
\tilde A_{ij}=\frac{1}{2\alpha}
\left(f_{ik} \hspace{-1mm}\zeroD_j \beta^k+f_{jk} \hspace{-1mm}
\zeroD_i \beta^k
-\frac{2}{3}f_{ij} \hspace{-1mm}\zeroD_k \beta^k\right)\,, \label{eq:Aij}
\eeqn
where indices of $\tilde A_{ij}$, $\tilde A^{ij}$, and $\zeroD_i$ are raised and lowered by $f^{ij}$ and $f_{ij}$. 
The condition $K=0=\pa_t K$ yields the equation for $\alpha$, which leads to the equation for $\chi:=\alpha \psi$ in the form,
\beqn
\zeroDelta \hspace{-1mm}\chi &=& \chi \psi^4\biggl[
  2\pi G\phi^{-1} (2S+\rho_{\rm h})
+\frac{7}{8}\tilde A_{ij} \tilde A^{ij}
\nonumber \\
&&~~~~~~~+\frac{1}{8}\omega \phi^{-2}\left\{
7\Pi^2 - (D_i \phi) D^i \phi \right\}
\nonumber \\
&&~~~~~~~-\frac{1}{4\alpha_{\rm s}^2 \phi^2}X \frac{\pa \hat K}{\pa X}
\left( 2(D_k\phi) D^k \phi +\Pi^2 \right) 
\nonumber \\
&&~~~~~~~+\frac{3}{4\phi} 
( D_i D^i \phi - 2\Box_g \phi)\biggr]\,,
\label{eq:trKJBD1}
\eeqn
where $S:=T_{ab}\gamma^{ab}$. 
Note that we replace $\Box_g \phi$ using
\beqn
\Box_g \phi&=&\frac{1}{F}\left(8\pi G \alpha_{\rm s}^2 T
-(\nabla^a X)(\nabla_a \phi) \frac{\pa F}{\pa X}
\right) \nonumber \\
&=&\frac{1}{F}\left( 8\pi G \alpha_{\rm s}^2 T
-\left\{ (D^k X)D_k\phi +(n^a\nabla_a X)\Pi \right\} \frac{\pa F}{\pa X}
\right)\,,\nonumber \\ \label{eq:boxg}
\eeqn
and will replace the Laplacian term of $D_i D^i\phi$ using the equation for $\phi$, as defined below.

For the scalar field, if we simply set $\Pi=0$, Eq.~\eqref{eqn:field_evol} (with $K=0$) leads to an elliptic equation for $\phi$,
\beqn
D_i D^i \phi &=& \psi^{-4}
\left[\zeroDelta \phi + 2\psi^{-1} (\zeroD_i \psi)\zeroD^{\,i} \phi
\right] \nonumber \\
&=&
- (D_i \ln \alpha)D^i\phi \nonumber \\
&& +F^{-1}\left[
  8\pi G \alpha_{\rm s}^2 T -(D_k\phi)(D^k X) \frac{\pa F}{\pa X}
  \right]\,, 
\label{eq:phi1}
\eeqn
with $X=(D_k\phi) D^k\phi/(16\pi G\alpha_{\rm s}^2 \phi^3)$. 
The treatment with $\Pi=0$ is justified in the case where the gravitational radiation reaction timescale is much longer than the orbital period, $2\pi/\Omega$.
With the choice of $\Pi=0$, the equation for $\Box_g \phi$ simplifies to
\beq \Box_g \phi=F^{-1}\left[ 8\pi
  G \alpha_{\rm s}^2 T -(D_k\phi)(D^k X) \frac{\pa F}{\pa X}\right]\,.
\eeq
Furthermore, Eqs.~\eqref{eq:Ham1}, \eqref{eq:Mom1}, and \eqref{eq:trKJBD1} are also simplified given the choice of $\Pi=0$.

To obtain the solution for spherical stars in exact equilibrium, we set $\Omega=0$, $\beta^k=0$, $\tilde A_{ij}=0$, and solve the elliptic equations only for $\psi$, $\chi$, and $\phi$ with appropriate boundary conditions at $r=0$ and $r\rightarrow \infty$.
The hydrostatic equation has the form,
\beq
\alpha h ={\rm const}. 
\eeq

The asymptotic behaviour of $\psi$, $\chi$, and $\phi$ for $r\rightarrow \infty$ is given by
\beqn
&&\psi \rightarrow 1 + \frac{M_{\rm ADM}}{2 r},\label{psi0}\\
&&\chi \rightarrow 1 - \frac{2M_{\rm K}-M_{\rm ADM}}{2 r},\label{chi0}\\
&&\phi \rightarrow 1 + \frac{2 M_{\rm S}}{r}\,,
\eeqn
where $M_{\rm ADM}$, $M_{\rm K}$, and $M_{\rm S}$ are the ADM mass, Komar mass, and scalar charge.
The tensor mass, which is a conserved quantity in scalar-tensor theories and the ADM mass in the Einstein frame, is defined from $M_{\rm ADM}$ and $M_{\rm S}$ by~\cite{Lee:1974clv}
\beqn
M_{\rm T}=M_{\rm ADM} + M_{\rm S}\,.
\eeqn
The virial relation, which is satisfied in stationary and quasi-equilibrium solutions, is written as~\cite{Shibata:2013vir}
\beq
M_{\rm K}=M_{\rm ADM} + 2M_{\rm S}=M_{\rm T} + M_{\rm S}\,.\label{virial}
\eeq

Equation~\eqref{eq:phi1} indicates that in the far zone, for which $X < \Lambda^4$ is satisfied, $|\phi-1|$ is of the same order of magnitude as $\alpha_{\rm s}^2 GM/r$, where $M$ denotes the mass of the system. 
Using the definition of $X$ in Eq.~\eqref{eqX}, the magnitude of $X/\Lambda^4$ is written as
\beq
\sim \frac{\alpha_{\rm s}^4 \lambda^4}{16\pi \ell_{\rm p}^2}
     \left(\frac{r_g}{r^2}\right)^2\,,
\eeq
where $r_g=GM/c^2$ is the gravitational radius.
Thus the screening effect occurs for $r \alt r_{\rm sc}:=\alpha_{\rm s}\lambda (r_g/\ell_{\rm p})^{1/2}$. 
Here $\lambda \approx 1.97 \times 10^{-11}\,{\rm cm}\, (\Lambda/1\,{\rm MeV})^{-1}$. 
In the following, we specify the strength of the screening by the dimensionless parameter
\beqn
\beta:=\frac{\lambda^8}{\ell_p^4 r_{g,\odot}^4}
&\approx& 1.20 \times 10^{28}
\left(\frac{\lambda}{5\times 10^{-11}\,{\rm cm}}\right)^8 \nonumber \\
&\approx& 1.08 \times 10^{28}
\left(\frac{\Lambda}{0.4\,{\rm MeV}}\right)^{-8},
\eeqn
where $r_{g,\odot}=GM_\odot/c^2$. 
Using this parameter, the radius of the screening region can be expressed as
\beqn
r_{\rm sc}&=&\alpha_{\rm s} \beta^{1/8} (r_g r_{g,\odot})^{1/2}
\nonumber \\
&=& 5.53\times 10^2 \,{\rm km} \left(\frac{\alpha_{\rm s}}{0.1}\right)
\left(\frac{\beta}{10^{28}}\right)^{1/8}
\left(\frac{r_g}{1.4r_{g,\odot}}\right)^{1/2}
\nonumber \\
&=& 5.58 \times 10^2\,{\rm km} \left(\frac{\alpha_{\rm s}}{0.1}\right)
\left(\frac{\Lambda}{0.4\,{\rm MeV}}\right)^{-1}
\left(\frac{r_g}{1.4r_{g,\odot}}\right)^{1/2}. \nonumber \\
\label{screenr}
\eeqn
Any object that has a screening radius larger than it's physical size would screen modifications to gravity within this region.

\begin{figure*}[t!]
\includegraphics[width=0.48\textwidth]{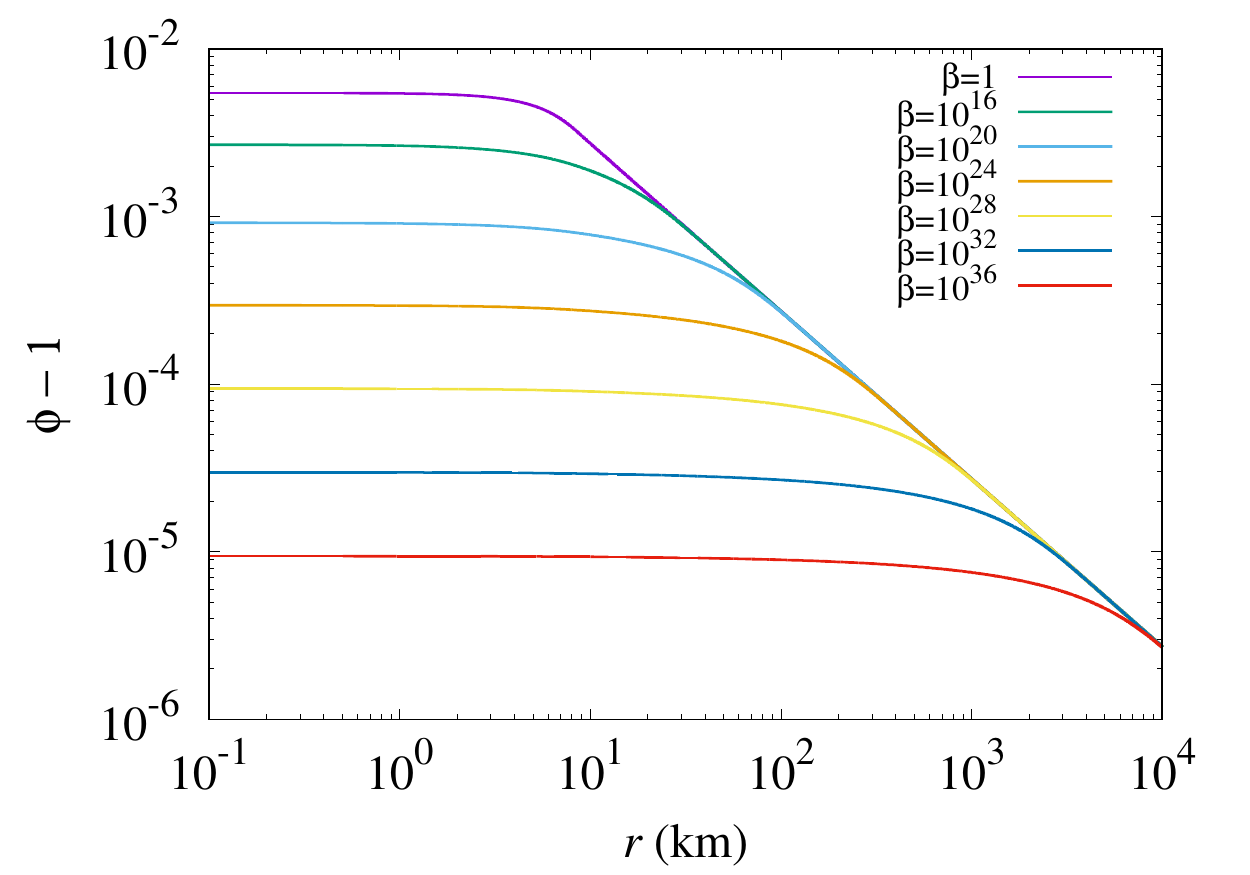}~~~
\includegraphics[width=0.48\textwidth]{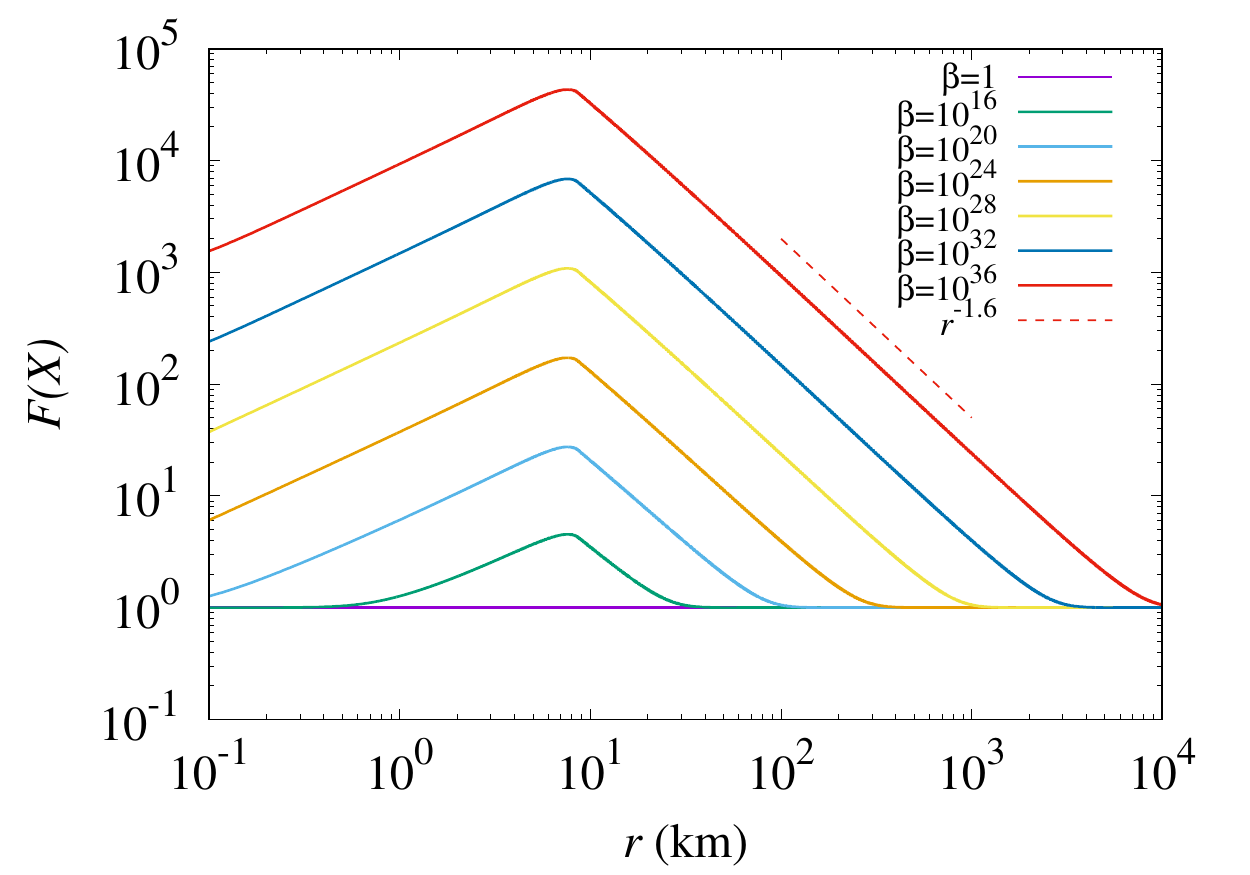}
\caption{
$\phi-1$ (left) and $F(X)$ (right) as functions of the radius in isotropic coordinates for spherical neutron stars of mass, $M_{\rm T}=1.4M_\odot$.
The dashed slope line in the right panel indicates that $F(X)$ outside the stellar surface is approximately proportional to $r^{-1.6}$.}
\label{fig:spher_NS}
\end{figure*}

\section{Spherical neutron stars}
\label{sec:spher_NS}
In this section we summarise how the screening effect appears in static spacetimes by showing solutions of spherical neutron stars of $M_{\rm T}=1.4M_\odot$ for a wide range of $\beta$, defined in Sec.~\ref{sec:initial_cond}.
We find that the qualitative behaviour of $\phi$, $F(X)$, and geometric quantities is essentially the same for other values of $M_{\rm T}$, and thus, we focus only on this specific mass case.
We fix $\alpha_{\rm s}=0.1$.
For the $1.4M_\odot$ neutron star, the stellar radius (circumferential radius) is $\approx 11.1$\,km and the scalar charge is $\approx 0.018M_\odot$ irrespective of the value of $\beta$.
The validity of the numerical equilibrium profile is confirmed by the fact that the virial relation is satisfied within a relative error $<10^{-4}$. 

Figure~\ref{fig:spher_NS} plots the profiles of $\phi-1$ (in the left panel) and $F(X)$ (right panel) as functions of the coordinate radius $r$ (in isotropic coordinates) for $\beta=1$ and $10^{16}$--$10^{36}$.
Note that for $\beta=1$, $F(X)\approx 1$ for the entire region, and hence, the solution may be considered as that in the FJBD theory. 
It is found that the central value of $\phi-1$, $\phi_c-1$, decreases with the increase of $\beta$, reflecting the screening effect.
The value of $\phi_c-1$ is approximately proportional to $\beta^{1/8}$, i.e., proportional to the screening radius, $r_{\rm sc}$, for $\beta \geq 10^{16}$.

The right panel of Fig.~\ref{fig:spher_NS} demonstrates that Eq.~\eqref{screenr} approximately indicates the screening region of $F(X) \agt 2$. For the larger values of $\beta$, we find a wider screening region, whereas for $\beta \alt 10^{16}$,
the screening region disappears.
Around the stellar centre, $F(X)$ approaches unity because $D_j \phi=0=\Pi$ in such a region, and thus, the screening is absent near the stellar centre. Note that the peak of $F(X)$ (and thus $X$ in our present choice) always appears near the stellar surface (which is located at $r \approx 8.9$\,km).
Outside the stellar surface, $F(X)$ decreases approximately proportional to $r^{-n}$, where $n \approx 1.6$ (denoted by the red dashed line on the plot). 
The reason for this is explained by the following analysis.
Outside the neutron star, Eq.~\eqref{eq:JBD2} is integrated to give (in the present case), 
\beqn
\alpha \psi^2  r^2 F  \pa_r \phi=8\pi G \alpha_{\rm s}^2
\int T \alpha \psi^6  r^2 dr=2M_{\rm T}\,.
\eeqn
Assuming that $F \propto r^{-n}$ and $\phi \propto r^{-p}$, the left-hand side is approximately proportional to $r^{1-p-n}$, resulting in $n=1-p$.
On the other hand, $X$ is approximately proportional to $(\pa_r \phi)^2 \propto r^{-2p-2}$, and for $X \gg 1$, $F(X) \propto X^2 \propto r^{-4p-4}$, resulting in $n=4p+4$. Thus we obtain $p=-3/5$ and $n=8/5$.
\footnote{We note that this relation should be satisfied sufficiently outside the matter source even for stationary and quasi-stationary spacetime (but the powers, $n$ and $p$, depend on the chosen function of $F(X)$) and that grasping the behaviour of $F(X)$ plays an important role for understanding the propagation property of scalar waves (see Sec.~\ref{sec:non-spher_NS}).}

For $X\ll1$, it scales as $X \propto r^{-4}$, and thus, $F(X)$ steeply approaches unity. 
Inside the stellar surface, $F(X)$ increases with the radius for $M_{\rm T}=1.4M_\odot$.
However, this is not always the case for high-mass neutron stars ($M_{\rm T}\agt 2M_\odot$ for the APR4 EOS), for which $T(=-\rho (1+\varep)+3P)$ can be positive for a very high-density region.
For such a star, $F(X)$ becomes unity not only at $r=0$ but also at an stellar interior; thus, $F(X)$ does not increase monotonically inside the star.
However, outside such a radius, $F(X)$ starts to increase again until the stellar surface.

\begin{figure}[t!]
\includegraphics[width=0.48\textwidth]{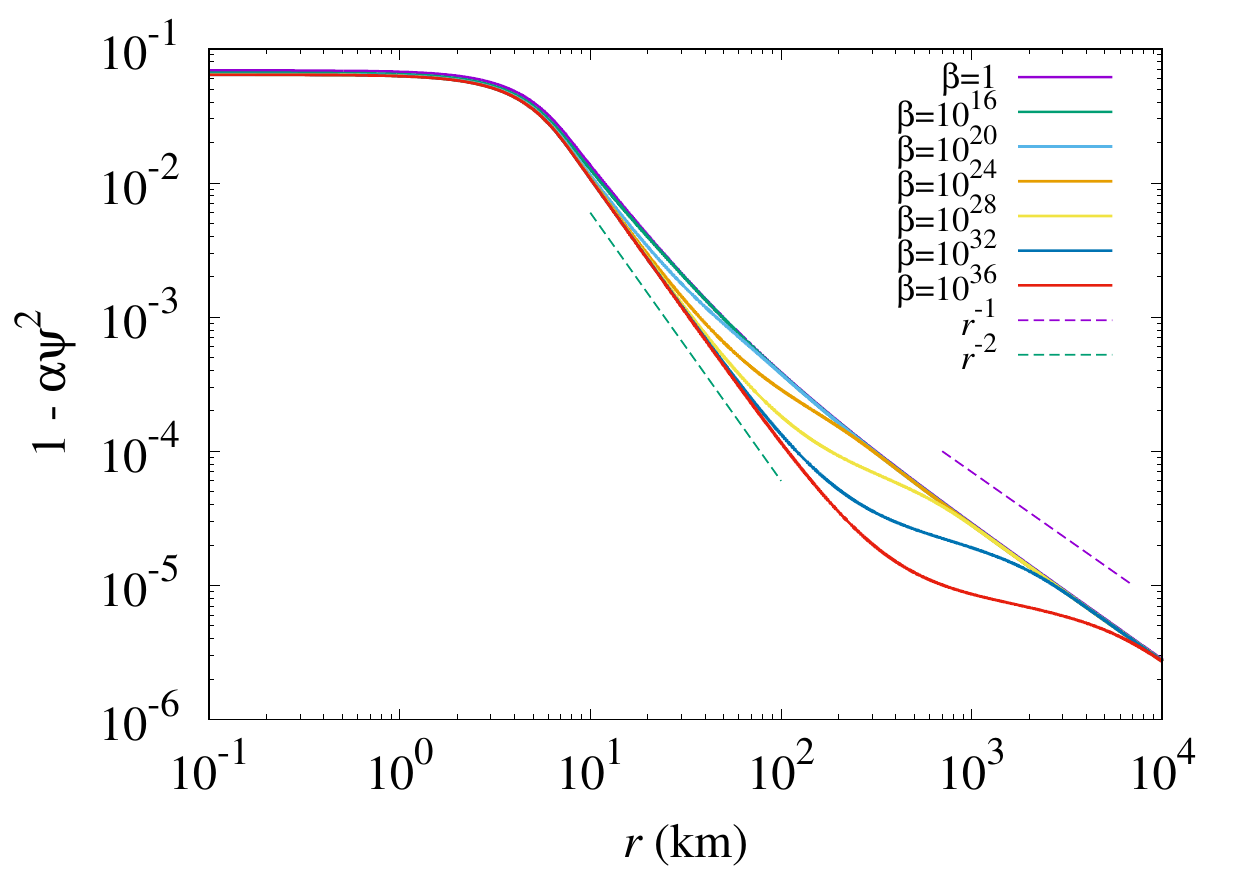}
\caption{$1-\alpha\psi^2$ as a function of the coordinate radius for spherical neutron stars of $M_{\rm T}=1.4M_\odot$.
The dashed lines denote the slope of $r^{-1}$ and $r^{-2}$.}
\label{fig:1-alpha_psi}
\end{figure}

Figure~\ref{fig:1-alpha_psi} plots $1-\alpha\psi^2=1-\chi\psi$ as a function of the coordinate radius, $r$.
In GR, where $M_{\rm S}=0$, this quantity falls off as $r^{-2}$ in isotropic coordinates (as shown by the green dashed line) due to the presence of the virial relation (see Eqs.~\eqref{psi0}, \eqref{chi0}, and \eqref{virial}). 
On the other hand, in the presence of the scalar charge, it goes as $M_{\rm S}/r$ (purple dashed line).
This plot shows that in the presence of screening, $1-\alpha\psi^2 \propto r^{-2}$, while outside the screening region it behaves approximately as $M_{\rm S}/r$. 
As already mentioned, the scalar charge depends only weakly on the value of $\beta$, and hence, in the far region, the profile of $\alpha\psi^2$ is essentially the same for any value of $\beta$~\footnote{We note that outside the screening region, the geometrical profile is the same as that in the FJBD theory with a Brans-Dicke parameter, as defined in Eq.~\eqref{omega}, $\omega=(-3+\alpha_{\rm s}^{-2})/2$, irrespective of the value of $\beta$.}. 

\section{Non-spherical oscillation of spherical neutron stars}
\label{sec:non-spher_NS}

\begin{figure}[t!]
\includegraphics[width=0.48\textwidth]{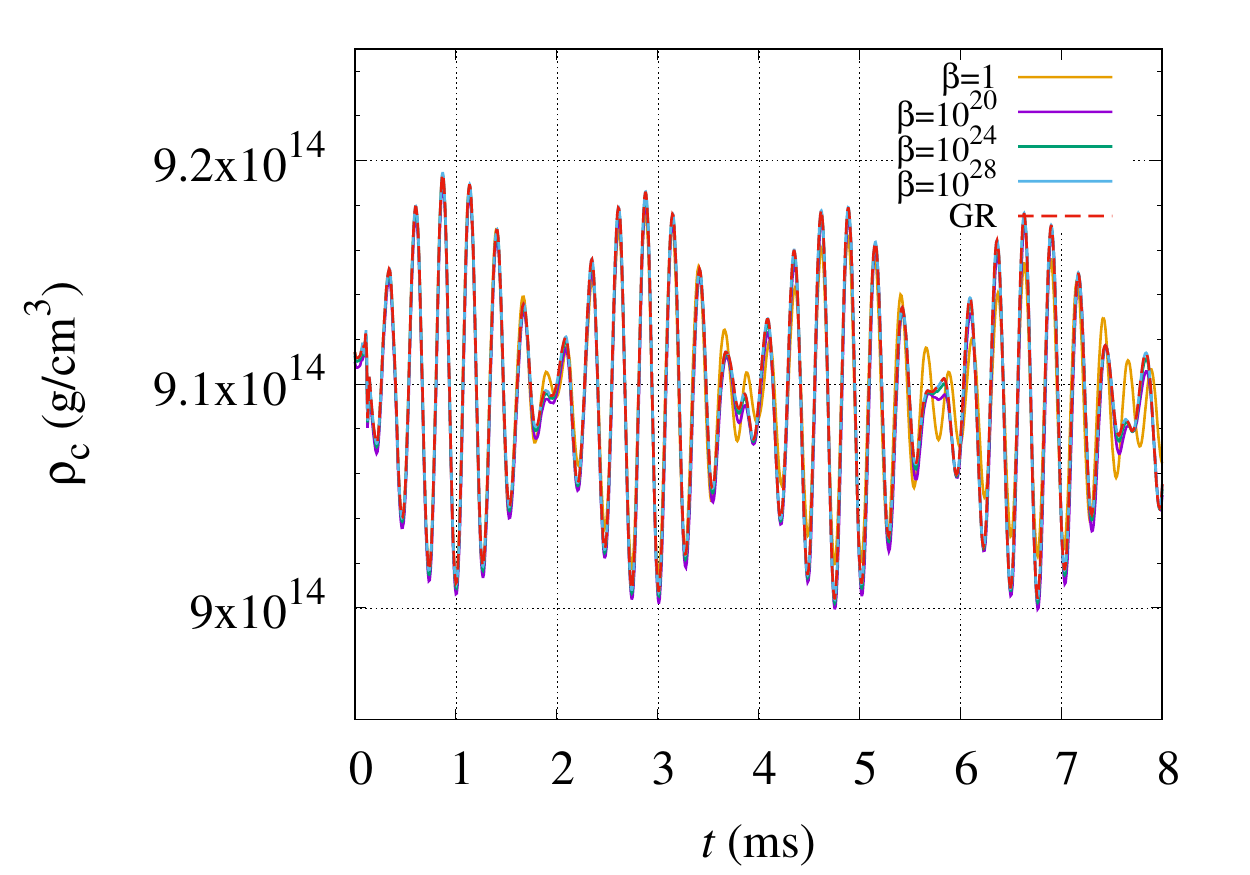}
\caption{Evolution of the central density for $\beta=1$, $10^{20}$,
  $10^{24}$, and $10^{28}$ as well as in GR.
  All the curves approximately overlap with each other.}
\label{fig:rho_c}
\end{figure}
Here, we explore the emission of scalar and gravitational waves from oscillating neutron stars~\footnote{We note that in scalar-tensor theory, Birkoff's theorem is not valid.}.
As a zeroth-order solution, we take the $M_{\rm T}=1.4M_\odot$ neutron stars from Sec.~\ref{sec:spher_NS}.
We also perform simulations for a high-mass neutron stars with $M_{\rm T}=1.9M_\odot$ and find very similar results to the $1.4M_\odot$ case.
Thus, in the following, we present only the results for $M_{\rm T}=1.4M_\odot$. 
All the simulations are performed for $\beta \leq 10^{32}$, i.e., $\Lambda \agt 0.1$\,MeV.

To excite a small quadrupole oscillation we superimpose
\beq
u_x=\sigma x~~~{\rm and}~~~u_y=-\sigma y\,,
\eeq
where we set $\sigma=1.0\times 10^3\,{\rm s}^{-1}$.
The oscillation velocity is at most 3\% of the speed of light near the stellar surface, and hence, the density and pressure profiles remain close to the spherical ones. 
However, the quadrupole mode, $l=|m|=2$, of scalar and gravitational waves is still appreciably excited so, in the following, we pay particular attention to this mode. 

The numerical simulations are performed using a fixed-mesh refinement code, {\tt SACRA}~\cite{Yamamoto:2008js}, covering the radius of spherical neutron stars by $N=45$ and 55 grid points in the finest computational domain.
We find that the dependence of the numerical results on the grid resolution is very weak in the present problem, and we always show the result for $N=55$ in the following.
For scalar waves we directly analyse $\phi-1$ in the far region of $r \agt \lambda_{\rm wave}$.
For gravitational waves, we extract the outgoing component of the complex Weyl scalar (the so-called $\Psi_4$).
For more details, see Appendix~\ref{app:extraction}. 
Our simulations are performed at longest for 15\,ms.
For high values of $\beta$, we find that it is in fact not trivial to perform a long-term simulation (with duration longer than 10\,ms) as a small numerical error often emerges in the primitive recovery process of determining $X$ from Eq.~(\ref{eqn:X2}) and in some cases leads to a pathological solution (see Appendix~\ref{app:3+1} for details).
However, it is still possible to draw an important conclusion from relatively short-term simulations as we will show. 
We leave developing an implementation for a long-term simulation (with duration of $\gg 10$\,ms) for future work.

We perform simulations for $\beta=1$, $10^{16}$, $10^{20}$, $10^{22}$, $10^{24}$, $10^{26}$, $10^{28}$, $10^{30}$, $10^{32}$, and $10^{36}$ (the corresponding $\Lambda$ for which are $\Lambda\sim \{1.28\times10^3,12.8,4.04,2.27,1.27,0.718,0.404,0.227,0.128,4.04\times10^{-2}\}$ MeV, respectively), as well as in GR (i.e., in the absence of the scalar field or $\phi=1$).
When $\beta=1$, $F(X) \approx 1$ in the entire region, and hence, the results are essentially the same as those in the FJBD theory.


Figure~\ref{fig:rho_c} shows the evolution of the central density for $\beta=1$, $10^{20}$, $10^{24}$, and $10^{28}$ as well as in GR. 
Due to the input perturbation, the star oscillates with time not only non-spherically but also spherically, and as a result, the central density also varies with time.
In this figure we can see that the oscillation pattern and amplitude depend very weakly on the value of $\beta$, although the ones with larger screening effect (i.e. $\beta \geq 10^{24}$) appear to agree best with GR.
Since the oscillation pattern is approximately identical for all the models, we may consider that the source of the scalar and gravitational wave emission is approximately identical in the present setting.

\begin{figure}[t!]
\includegraphics[width=0.48\textwidth]{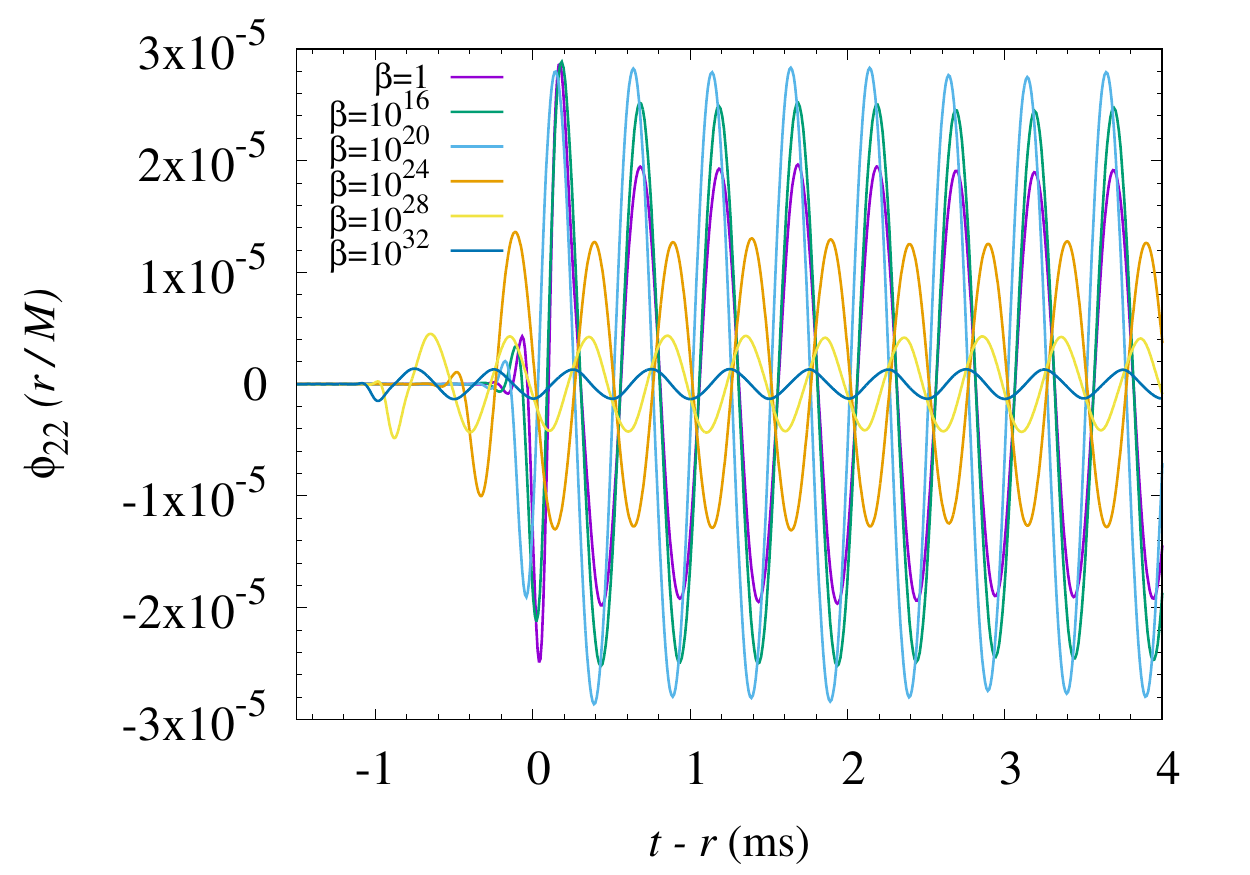}
\caption{Waveforms of the quadrupole mode for scalar waves as functions of $t-r$ for $\beta=1$, $10^{16}$, $10^{20}$, $10^{24}$, $10^{28}$, and $10^{32}$. The waveforms extracted at $r=591$\,km are shown together. For $\beta=10^{28}$ and $10^{32}$, correction factors of $F^{0.6}$ and $F^{0.5}$ are multiplied, respectively (see Appendix~\ref{app:extraction} on the correction factor). 
\label{fig:22pole_waveform}}
\end{figure}

\begin{figure}[t]
\includegraphics[width=0.48\textwidth]{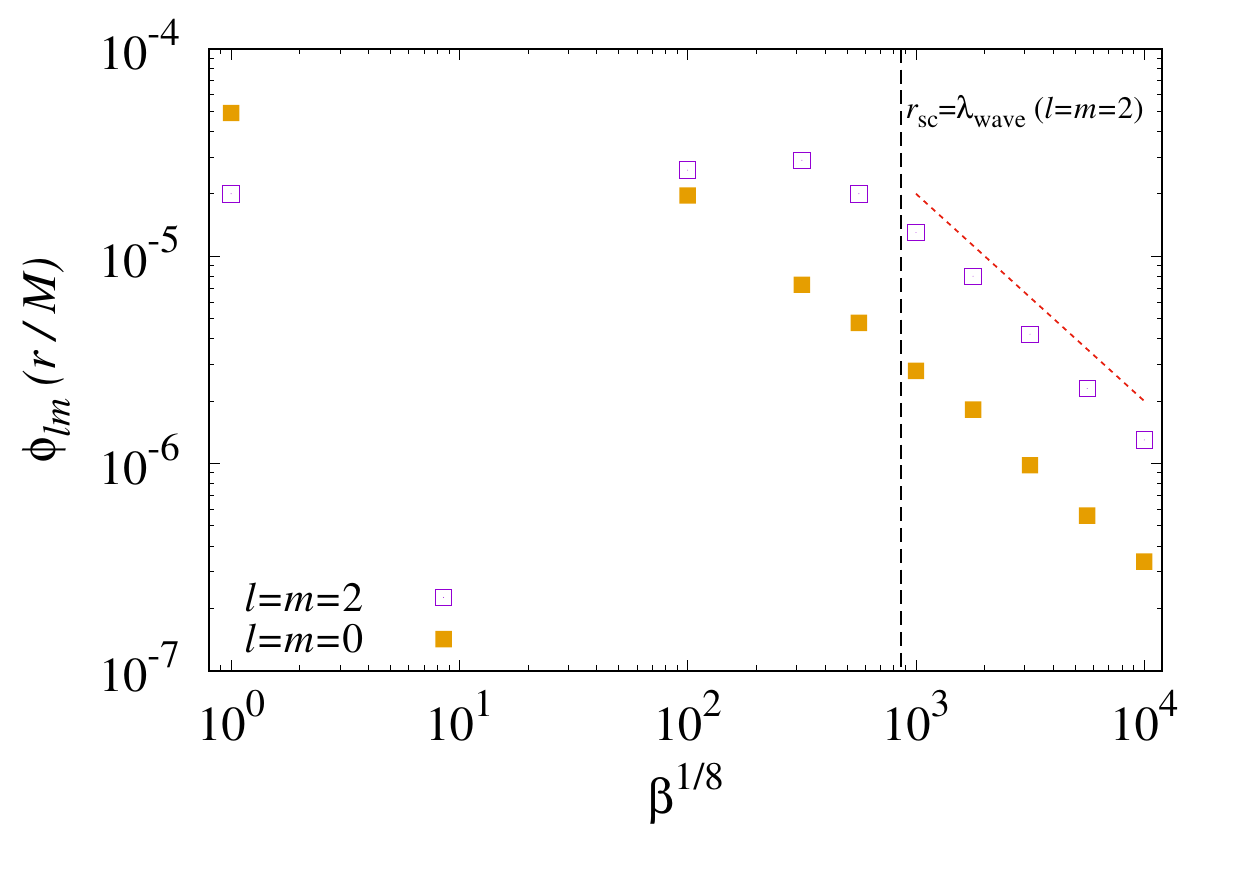}
\caption{Asymptotic amplitudes of scalar waves as functions of $\beta^{1/8} (\propto r_{\rm sc})$ for the quadrupole ($l=m=2$) mode (hollow squares) and for the monopole ($l=m=0$) mode (filled squares). 
For the monopole mode the asymptotic amplitude of $\pa_t \phi_{00} r_{\rm ex}$ is plotted.
The red dotted line denotes $\propto \beta^{-1/8}$, which indicates that the asymptotic amplitude of scalar waves decreases approximately as $r_{\rm sc}^{-1}$ for the parameter space of $r_{\rm sc} > \lambda_{\rm wave}$ irrespective of the modes considered.
The black vertical dashed line denotes the value of $\beta$ which satisfies $r_{\rm sc}=\lambda_{\rm wave}$ for the quadrupole mode.
For the monopole mode, $r_{\rm sc}=\lambda_{\rm wave}$ is satisfied at $\beta^{1/8} \sim 460$ in the present case.
\label{fig:scalar_amplitude}}
\end{figure}


We indeed find that the gravitational waveforms depend only very weakly on the value of the $\beta$ parameter (see Fig.~\ref{fig:extraction} in Appendix~\ref{app:extraction}).
In particular, for $\beta \agt 10^{24}$, i.e., where the screening effect to the scalar wave generation becomes noticeable, the gravitational waveforms are in a good agreement with those in GR (although about 10\% level disagreement is found irrespective of $\beta$ values presumably due to the numerical error).
For $\beta=1$ (approximately same as the FJBD case), the amplitude of gravitational waves is slightly higher than those in GR, reflecting a significant contribution of the scalar field in determining the stellar profile.

By contrast, the amplitude of scalar waves depends strongly on the $\beta$ parameter in spite of approximately the same emission source, although the frequency is always identical in all cases. 
Figure~\ref{fig:22pole_waveform} shows the quadrupole mode of scalar waves as a function of $t-r$ for $\beta=1$, $10^{16}$, $10^{20}$, $10^{24}$, and $10^{28}$, and Fig.~\ref{fig:scalar_amplitude} summarises the wave amplitude as a function of $\beta^{1/8}$ (see the hollow squares). 
These plots show that, for $r_{\rm sc} < \lambda_{\rm wave}$, scalar waves are emitted to the far zone broadly in the same manner as in the FJBD case, in which screening is absent. 
Interestingly, we find that for $r_{\rm sc}> \lambda_{\rm wave}$, where the screening effect plays an important role, the amplitude of scalar waves is suppressed\footnote{Besides the amplitude dependence on $\beta$, a phase misalignment among the scalar waves is found. The reasons for this are discussed in more detail in Appendix~\ref{app:extraction}.}. 
The reason for the suppression can be seen in the large value of $F(X)$ inside the screening radius.
By rewriting Eq.~\eqref{eq:JBD2} as
\beq
\nabla^a \nabla_a \phi + (\nabla^a \ln F) \nabla_a \phi
= 8\pi G \alpha_{\rm s}^2 T F^{-1}\,, \label{eq:JBD2a}
\eeq
we can see that the factor $F^{-1}$ suppresses the scalar wave generation associated with the matter motion by $T$.

One point to be added is that the suppression in the wave amplitude is not as large as the one by the $F$ factor. 
For example, for $\beta=10^{28}$, $F>10^2$ for $r=1$--20\,km, while the suppression fraction in the wave amplitude is $\sim 1/10$. 
The reason for this is that the wave amplitude, defined by $\phi_{22}(r/M)$, increases during the outward propagation inside the screening radius, i.e., for $r < r_{\rm sc}$, by $F^{-\eta}$ (see Appendix~\ref{app:extraction} for details).

As we can see in Fig.~\ref{fig:scalar_amplitude}, the amplitude of quadrupole scalar waves depends only weakly on $\beta$ for $r_{\rm sc} \alt \lambda_{\rm wave}/3$ (i.e., $\beta \alt 10^{20}$), with the steep decline starting only at $r_{\rm sc}\sim \lambda_{\rm wave}$
This suggests that the suppression effect by $F^{-1}$ in the wave generation and the amplification effect during the propagation of waves in the region of $F>1$ is likely to be balanced for the quadrupole mode.

\begin{figure}[t]
\includegraphics[width=0.48\textwidth]{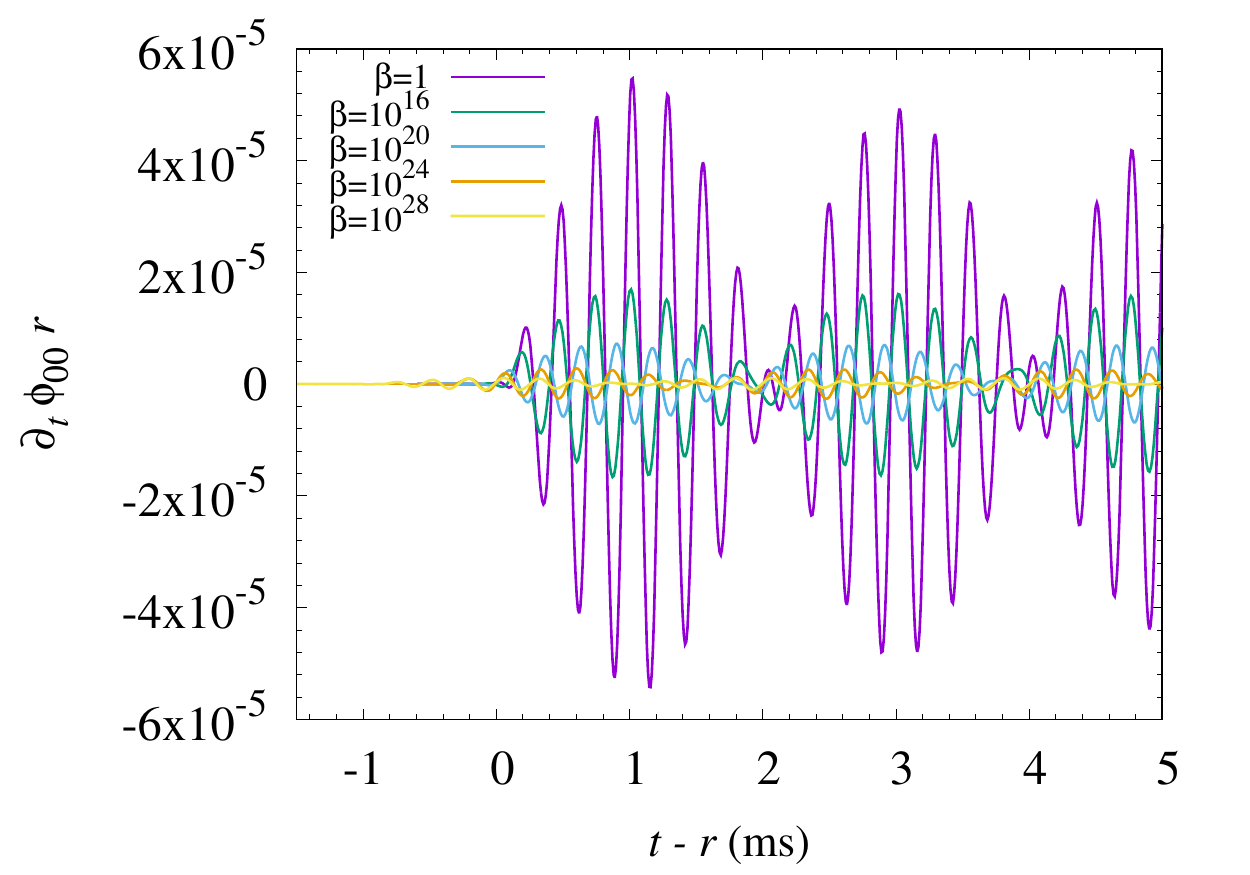}
\caption{Waveforms ($\pa_t \phi_{00} r$) of the monopole mode for scalar waves as functions of $t-r$ for $\beta=1$, $10^{16}$, $10^{20}$, $10^{24}$, and $10^{28}$. The waveforms extracted at $r=591$\,km are shown together. 
\label{fig:monopole_waveform}}
\end{figure}


On the contrary, for the monopole we find that the steep decline does not start at the point of $r_{\rm sc}=\lambda_{\rm wave}$, where the wavelength of the monopole mode is $\sim 80$\,km, and thus, $r_{\rm sc}=\lambda_{\rm wave} (l=m=0)$ is satisfied at $\beta \approx 2 \times 10^{21}$. 
The decrease of the asymptotic amplitude is again approximately proportional to $\beta^{-1/8}$, satisfied for a wide range of $\beta$ values, as can be seen in the filled squares of Fig.~\ref{fig:scalar_amplitude}.
This can also be seen clearly in Fig.~\ref{fig:monopole_waveform}, which shows the monopole waveforms for $\beta=1$, $10^{16}$, $10^{20}$, $10^{24}$, and $10^{28}$, extracted at $r=591$\,km.
(We should note that in this case, we analyse $\pa_t \phi_{00}$ simply because it is clearer to see the oscillation mode.)
This feature is in agreement with the results found in Ref.~\cite{Bezares:2021yek}, in which the authors analyse the amplitude of $l=m=0$ scalar waves emitted by the spherical oscillation of a neutron star. 
Therefore, we can conclude that while the steep decline of the amplitude is always found irrespective of the modes for $r_{\rm sc} > \lambda_{\rm wave}$, in the case of $r_{\rm sc} \alt \lambda_{\rm wave}$, the emergence of the screening effect on the scalar wave emission depends on the modes considered, presumably reflecting the generation mechanism (e.g., the main generation location) of each mode.

Finally, we consider the results of Ref.~\cite{Bezares:2021dma}, in which the authors explored scalar and gravitational waves from the late inspiral phase of binary neutron stars, in the case where $\lambda_{\rm wave} (\agt 300\,{\rm km}) > r_{\rm sc} \approx 140$\,km\footnote{Note that for typical binary neutron stars, the orbital period at their innermost stable circular orbits is $\sim 2$\,ms, and thus, the wavelength of the quadrupole mode is $\agt 300$\,km.}.
From our present analysis our suspicion is that one cannot expect the screening effect to appear in the quadrupole mode during the inspiral in that setting.
We argue that to fully understand the screening effect on the quadrupole mode, one should consider parameters, for which $r_{\rm sc} > 300$\,km. 
As we discussed above, the screening effect may appear in the low-multipole mode even for the case of $r_{\rm sc} < \lambda_{\rm wave}$, which means one can expect to find screening in the dipole mode even for small screening radii, as they report.

\section{Discussion}
\label{sec:discussion}

By analysing an oscillating spherical neutron star, we have confirmed that in a scalar-tensor theory with kinetic screening, the scalar wave emission is suppressed for a screening radius, $r_{\rm sc}$, larger than the wavelength of the emitted waves, $\lambda_{\rm wave}$, irrespective of multipole modes considered.
Therefore, inside the screening radius satisfying the condition $r_{\rm sc} >\lambda_{\rm wave}$, both the matter motion and wave emission are essentially the same as those in GR. 

However, for a screening radius $r_{\rm sc} \alt \lambda_{\rm wave}$, we have found emission of quadrupole scalar waves with a large amplitude, comparable to that in FJBD theory and additionally that the amplitude depends only weakly on $r_{\rm sc}$.
Therefore, if the analysis were to be restricted to small values of $r_{\rm sc}$, this could have lead to the conclusion that no screening effect is present in these theories.
We argue that to fully understand the nature of this theory it is necessary to perform the analysis at a wide range of $r_{\rm sc}$ values, including $r_{\rm sc} > \lambda_{\rm wave}$.

For the monopole mode, we have confirmed that the screening effect appears even for the case of $r_{\rm sc}<\lambda_{\rm wave}$ as was also found previously in Ref.~\cite{Bezares:2021yek}. 
Furthermore, we have found that, irrespective of the modes considered, the asymptotic scalar wave amplitude decreases roughly as $r_{\rm sc}^{-1}$ when $r_{\rm sc} > \lambda_{\rm wave}$.
For ground-based gravitational wave detectors, such as advanced LIGO and advanced Virgo, the lower limit of the frequency in the sensitive band of gravitational waves is about 10\,Hz, and thus, the upper limit of the observable wavelength is $\approx 3\times 10^4$\,km.
Therefore, if $r_{\rm sc} > 3 \times 10^4$\,km, it would be difficult to detect scalar-type gravitational waves due to the screening in this kind of scalar-tensor theories.
A number of previous solar system experiments have reported no evidence for the presence of a scalar field effect, which implies that $r_{\rm sc}$ has to be larger than the solar radius ($\approx 7 \times 10^5$\,km). 
Thus, the detection of scalar waves, e.g., from neutron-star oscillations and inspiraling binary neutron stars, by the ground-based gravitational wave detectors might be unlikely in kinetic screening theories.\footnote{Note, however, that when a black hole is formed dynamically, scalar waves of a characteristic wave shape with an appreciable amplitude can be emitted even in the presence of the screening effect irrespective of $r_{\rm sc}$ because the non-uniform scalar field disappears after the formation of the black hole~(e.g., Refs.~\cite{1972CMaPh..25..167H,Shibata:1994qd,Bezares:2021yek}).}

Our analysis in this paper has focused only on scalar and gravitational waves from oscillating neutron stars. To fully understand the emission mechanism of scalar waves in screened modified gravity theories, we should also perform simulations for other systems, such as binary neutron stars for a wide range of $r_{\rm sc}$.
As we have pointed out here, the emissivity of scalar waves is determined by the profile of $F(X)$, and if the profile for other systems is similar to that of single neutron stars, we can expect the conclusion to be the same; 
i.e., that the scalar wave emission is suppressed in the presence of screening with $r_{\rm sc} > \lambda_{\rm wave}$ irrespective of the multipole modes.
Thus, the question is what the profile of $F(X)$ is for other systems. We leave this further investigation for subsequent work.

\acknowledgments

We thank Kyohei Kawaguchi, Shinji Mukohyama, Miguel Bezares, Enrico Barausse Marco Crisostomi, and Carlos Palenzuela for helpful discussions and useful comments on this draft. This work was in part supported by Grant-in-Aid for Scientific Research (Grant No.~JP20H00158) of Japanese MEXT/JSPS. Numerical computations were performed on Sakura at Max Planck Computing and Data Facility.

\appendix

\section{3+1 formulation}\label{app:3+1}


Here we describe the 3+1 form of the gravitational and scalar fields equations.
By contracting Eq.~\eqref{eq:JBD1} with $n^a n^b$, the Hamiltonian constraint is derived as
\beqn
&&R_k^{~k} + K^2 -K_{ij}K^{ij}\nonumber \\
&&=16\pi G \phi^{-1} \rho_{\rm h}
+ \frac{\omega}{\phi^{2}} \left[\Pi^2 + (D_i\phi)D^i\phi\right]
\nonumber \\
&&- \frac{2 \Pi^2}{\alpha_{\rm s}^2\phi^2}X\frac{\pa \hat K}{\pa X}
+\frac{2}{\phi}(-K \Pi + D_i D^i \phi)\,,
\label{eq:Ham}
\eeqn
where $R_k^{~k}$ is the three-dimensional Ricci scalar.  

And contracting Eq.~\eqref{eq:JBD1} with $n^a \gamma^b_{~i}$ gives the momentum constraint,
\beqn
D_i K^i_{~j} - D_j K&=&8\pi G \phi^{-1} J_j 
\nonumber \\
&+&\left(\omega-\frac{X}{\alpha_{\rm s}^2}\frac{\pa \hat K}{\pa X}\right)
\phi^{-2}\,\Pi \,D_j\phi \nonumber \\
&+&\phi^{-1}(D_j \Pi - K^i_{~j} D_i \phi)\,,
\eeqn
and so the evolution equation can be obtained by contracting Eq.~\eqref{eq:JBD1} with $\gamma^a_{~i} \gamma^b_{~j}$,
\beqn
\pa_t K_{ij} &=&\alpha R_{ij}-8\pi G \alpha \phi^{-1} 
\left[S_{ij}-\frac{1}{2}\gamma_{ij}(S-\rho_{\rm h})\right] \nonumber \\
&+&\alpha(-2K_{ik} K_j^{~k}+K K_{ij}) \nonumber \\
&-&D_i D_j \alpha +\beta^k D_k K_{ij}+K_{ik} D_j \beta^k+K_{jk} D_i \beta^k 
\nonumber \\
&-&\alpha \left(\omega-\frac{X}{\alpha_{\rm s}^2}\frac{\pa \hat K}{\pa X}\right)
\phi^{-2} (D_i \phi) D_j \phi \nonumber \\
&-&\alpha \phi^{-1} \left(
D_i D_j\phi-K_{ij}\Pi \right)-\frac{\alpha}{2\phi}\gamma_{ij}\Box_g \phi
\nonumber \\
&-&\frac{\alpha}{2\phi^2}\gamma_{ij}\left((D_k\phi) D^k\phi-\Pi^2\right)
\frac{X}{\alpha_{\rm s}^2}\frac{\pa \hat K}{\pa X}\,, 
\label{eq:JBD9}
\eeqn
where $R_{ij}$ is the spatial Ricci tensor and $S_{ij}:=T_{ab}\gamma^a_{~i}\gamma^b_{~j}$ with $S$ its trace.

Equation~\eqref{eq:JBD9} together with the Hamiltonian constraint yields the following evolution equation for $K$,
\beqn
(\pa_t - \beta^k \pa_k) K 
&=&4\pi G \alpha \phi^{-1} (S+\rho_{\rm h})+\alpha K_{ij} K^{ij}-D_i D^i \alpha
\nonumber \\
&+&\alpha \omega \phi^{-2} \Pi^2 + \alpha \phi^{-1}
\left( D_i D^i \phi - K \Pi \right)\nonumber \\
&-&\frac{\alpha X}{2\alpha_{\rm s}^2 \phi^2} \frac{\pa \hat K}{\pa X}
\left( (D_k\phi) D^k \phi + \Pi^2\right)
-\frac{3\alpha}{2 \phi}\Box_g\phi\,,\nonumber \\
\label{eq:trKJBD}
\eeqn
and thus, the evolution equation for $\tilde A_{ij}=\psi^{-4} (K_{ij}-K\gamma_{ij}/3)$, where $\psi=({\rm det}\,\gamma_{ij})^{1/12}$ is the conformal factor, is written in the form
\beqn
&&(\pa_t-\beta^k\pa_k) \tilde A_{ij} =\frac{\alpha}{\psi^{4}}
\left(R_{ij}-\frac{1}{3}\gamma_{ij}R_k^{~k}\right) \nonumber \\
&&~~~~~-\psi^{-4}\left(
D_i D_j \alpha - \frac{1}{3}\gamma_{ij} D_k D^k \alpha \right)
\nonumber \\
&&~~~~~  +\tilde A_{ik} \pa_j \beta^k +\tilde A_{jk} \pa_i \beta^k
  -\frac{2}{3}\tilde A_{ij} \pa_k \beta^k
\nonumber \\
&&~~~~~+\alpha \left(K \tilde A_{ij} -2 \tilde A_{ik} \tilde A_j^{~~k}\right)
-8\pi G \frac{\alpha}{\psi^{4} \phi} 
\left(S_{ij}-\frac{1}{3}\gamma_{ij}S\right) \nonumber \\
&&~~~~~-\frac{\alpha}{\psi^{4}\phi^2} \left(\omega-\frac{X}{\alpha_{\rm s}^2}
     \frac{\pa \hat K}{\pa X}\right) \left[ (D_i \phi) D_j \phi
  -\frac{1}{3}\gamma_{ij}(D_k \phi) D^k \phi \right] \nonumber \\
&&~~~~~-\frac{\alpha}{\psi^{4} \phi}
\left(D_i D_j\phi-\frac{1}{3}\gamma_{ij} D_k D^k \phi
-\psi^4 \tilde A_{ij}\Pi \right)\,. 
\label{eq:JBD10}
\eeqn

The term, $\Box_g \phi$, in the right-hand side of Eq.~\eqref{eq:trKJBD} is undesirable in numerical evolution because of the presence of the time derivative of $\Pi$.
Therefore, to handle this term, we use the following expression of $\Box_g\phi$,
\beqn
\Box_g \phi 
&=& D_k D^k \phi + \frac{D_k \alpha}{\alpha} D^k \phi -K \Pi
+ n^a \nabla_a \Pi \nonumber \\
&=& D_k D^k \phi + \frac{D_k \alpha}{\alpha} D^k \phi -K \Pi
\nonumber \\
&& + \frac{\phi}{\alpha} (\pa_t-\beta^k\pa_k)
\left(\frac{\Pi}{\phi}\right)
-\frac{\Pi^2}{\phi}\,, \label{eq:JBD4}
\eeqn
and redefine the evolution equation for $\bar K:=K + 3\Pi/(2\phi)$ as
\beqn
&&(\pa_t - \beta^k \pa_k) \bar K 
=4\pi G \alpha \phi^{-1} (S+\rho_{\rm h})+\alpha K_{ij} K^{ij}-D_i D^i \alpha
\nonumber \\
&&\hskip 1.2cm +\alpha \left(\omega+\frac{3}{2}\right)
\phi^{-2} \Pi^2 -\frac{1}{2} \alpha \phi^{-1}
\left( D_i D^i \phi - K \Pi \right)\nonumber \\
&&\hskip 1.2cm
-\frac{\alpha X}{2\alpha_{\rm s}^2 \phi^2} \frac{\pa \hat K}{\pa X}
\left( (D_k\phi) D^k \phi + \Pi^2\right)
-\frac{3}{2 \phi}(D_k\alpha) D^k\phi\,,\nonumber \\
\label{eq:trKJBD2}
\eeqn
which guarantees the hyperbolicity of the geometric equations.

Equation~\eqref{eq:JBD2} is rewritten into a set of equations, (\ref{eq:JBD5a}) and (\ref{eqn:field_evol}), which are first-order in the time derivatives. 
Once $\phi$ and $\hat\Pi(=F(X)\Pi)$ are determined from these equations, $X$ (as well as $F(X)$ and $\Pi$) are obtained from Eq.~\eqref{eqX}, which is considered to be an algebraic equation for $X$ (see Eq.~\eqref{eqn:X2}).
For the present choice of $\hat K(X)$ (and $F(X)$), Eq.~\eqref{eqn:X2} has one or two or three real solutions for $X$.
For a small value of $\hat \Pi^2$, there is only one real solution.
However, for a value of $\hat \Pi^2$ larger than a critical value, there are more than two real solutions.
For the case that there are two real solutions, one should be a multiple solution.
In this case, the solution satisfies not only Eq.~\eqref{eqn:X2} but also the following,
\beq
\frac{df(X)}{dX}=1-\frac{\hat\Pi^2}{8\pi G \alpha_{\rm s}^2 \phi^3 F(X)^3}
     \frac{dF}{dX}=0\,.
\label{eqX22}
\eeq
This solution ($df(X)/dX=0$) has a pathology, and hence, in its presence the computation breaks down (see below).
Therefore, for a problem in which $\hat\Pi$ is initially small everywhere (i.e., $f(X) > 0$), but later it increases significantly leading to $f(X) \leq 0$ at points, it is possible that the computation breaks down.

If $X$ is determined, $K$ is obtained from $K=\bar K-3\Pi/(2\phi)$.
We also note that $D_j X$, which appears in the computation of $D_j F=(dF/dX)D_j X$, is calculated as
\beqn
D_j X&=&-\frac{3X}{\phi} D_j \phi
+\frac{1}{8\pi G \alpha_{\rm s}^2 \phi^3}\biggl[
    (D_j D_k\phi) D^k\phi - \frac{\hat\Pi \,D_j\hat\Pi}{F(X)^2}
  \nonumber \\
&& +\frac{\hat\Pi^2}{F(X)^3}\frac{dF}{dX} D_j X \biggr]\,, \label{eqX3}
\eeqn
and hence,
\beqn
D_j X&=&\left[-\frac{3X}{\phi} D_j \phi
+\frac{1}{8\pi G \alpha_{\rm s}^2 \phi^3}\biggl\{
(D_j D_k\phi) D^k\phi - \frac{\hat\Pi \,D_j\hat\Pi}{F(X)^2}\biggr\}
\right]
  \nonumber \\
  && \times
  \left[1-\frac{1}{8\pi G \alpha_{\rm s}^2 \phi^3}
    \frac{\hat\Pi^2}{F(X)^3}\frac{dF}{dX} \right]^{-1}\,. \label{eqX4}
\eeqn
This shows that $1-\hat\Pi^2/(8\pi G \alpha_{\rm s}^2 \phi^3F^3)(dF/dX)$ (i.e., $df/dX$) has to be non zero in general.
This implies that if the solution of $X$ is a multiple root of Eq.~\eqref{eqn:X2}, a discontinuity appears in the scalar field and the computation in general breaks down in the present formulation. 
In this work, we present results for which such a pathology is not encountered.

During the numerical simulation, we examine the violation of the Hamiltonian constraint by monitoring the following quantity:
\beqn
\langle H \rangle=
{1 \over M_*}\int {|H| \over \sum_l |H_l|} \rho_* d^3x,
\eeqn
where $H$ is defined by the left-hand side minus the right-hand side of Eq.~(\ref{eq:Ham}), 
$H_l$ denotes each individual term in Eq.~(\ref{eq:Ham}) so that $H=\sum_l H_l$, 
and $M_*$ is the rest mass of the system defined by
\beqn
M_*=\int \rho_* d^3x. 
\eeqn
We find that $\langle H\rangle$ remains to be always of order $10^{-4}$ during the simulation time in our present grid resolution if the simulation is successful; no indication of the growth of the constraint violation is found. For higher values of $\beta$, the magnitude of $\langle H \rangle$ is larger; e.g., for $\beta=10^{32}$ it is $\alt 10^{-3}$. For $\beta > 10^{32}$ with which stable evolution is not successful, it can quickly grow when the code crashes. This suggests that for such cases, higher grid resolution might be necessary for the successful simulation.

\section{Extraction method}\label{app:extraction}
\begin{figure*}[t]
\includegraphics[width=0.46\textwidth]{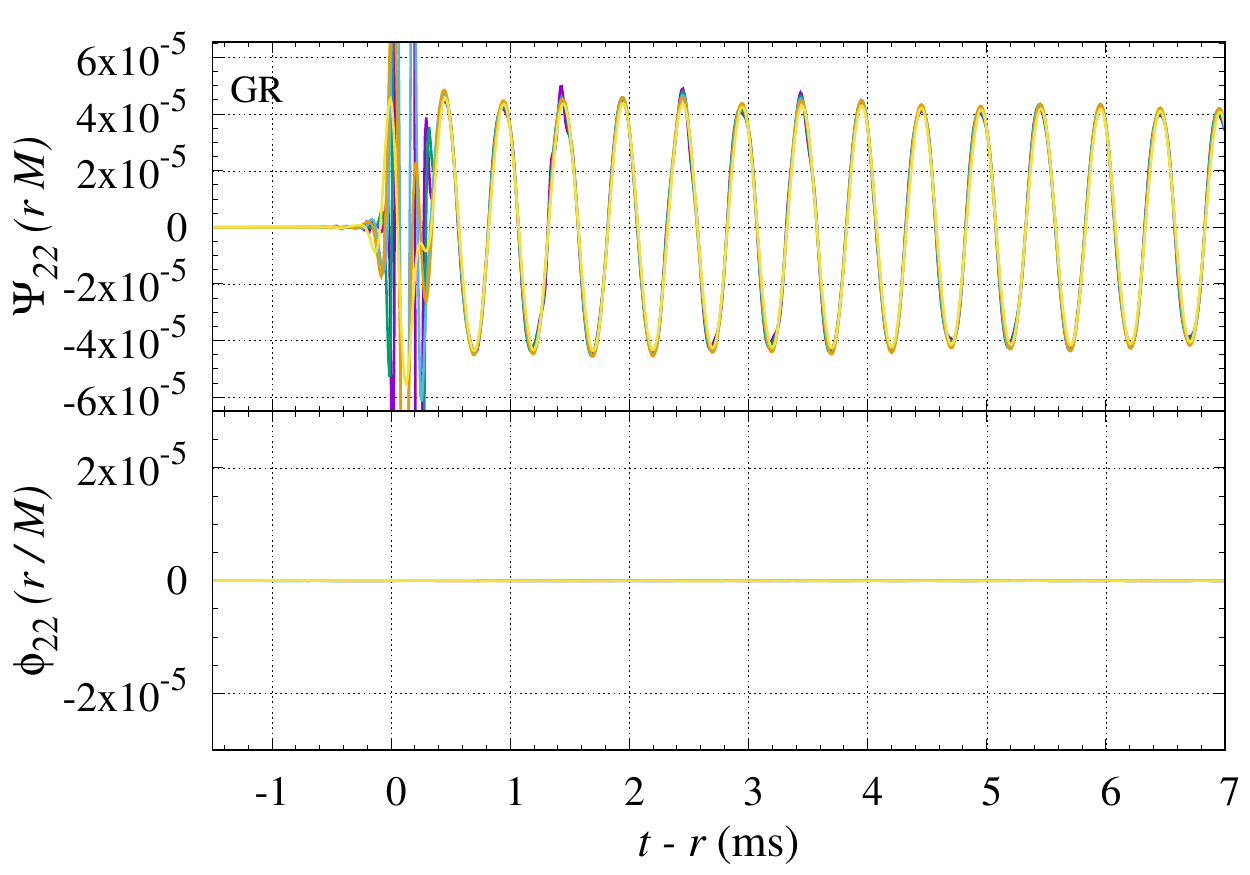}~~~
\includegraphics[width=0.46\textwidth]{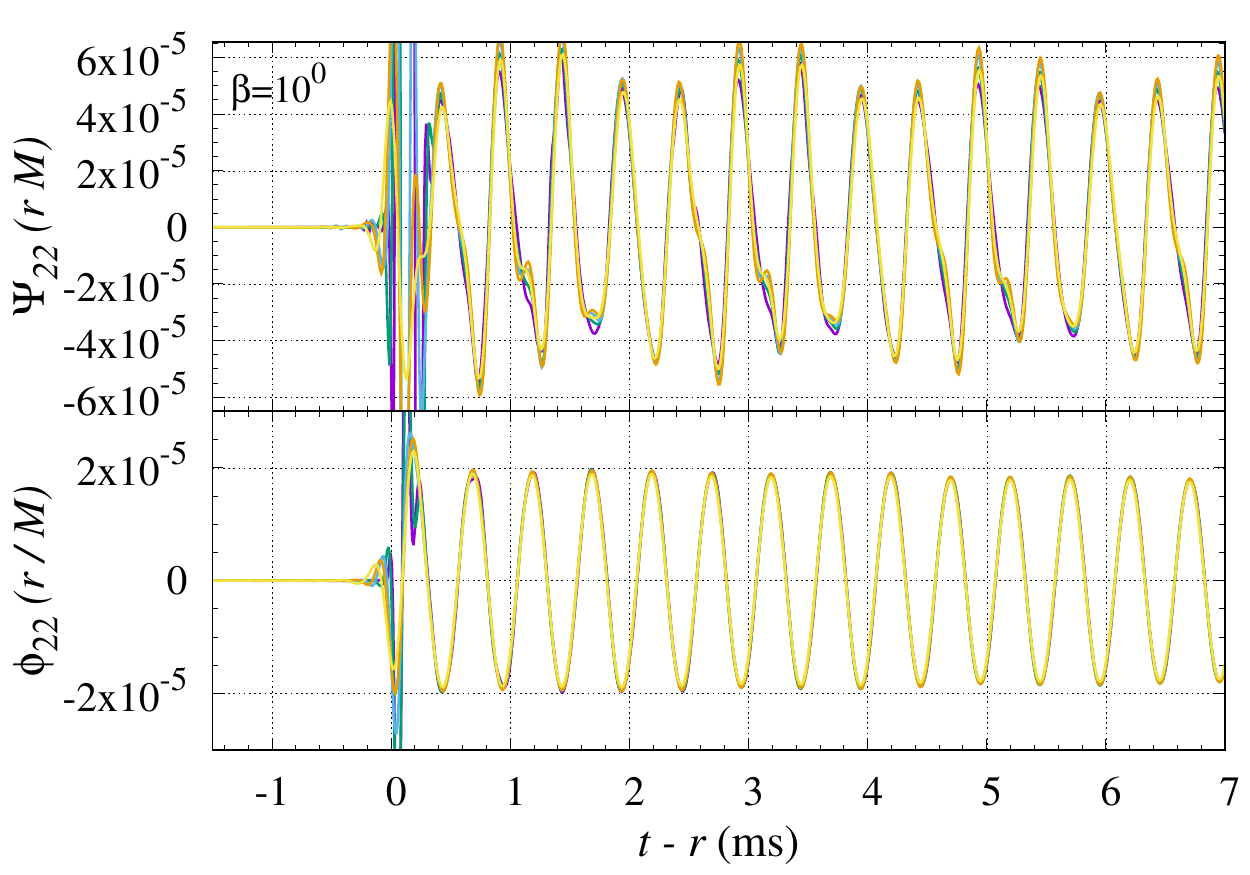}\\
\includegraphics[width=0.46\textwidth]{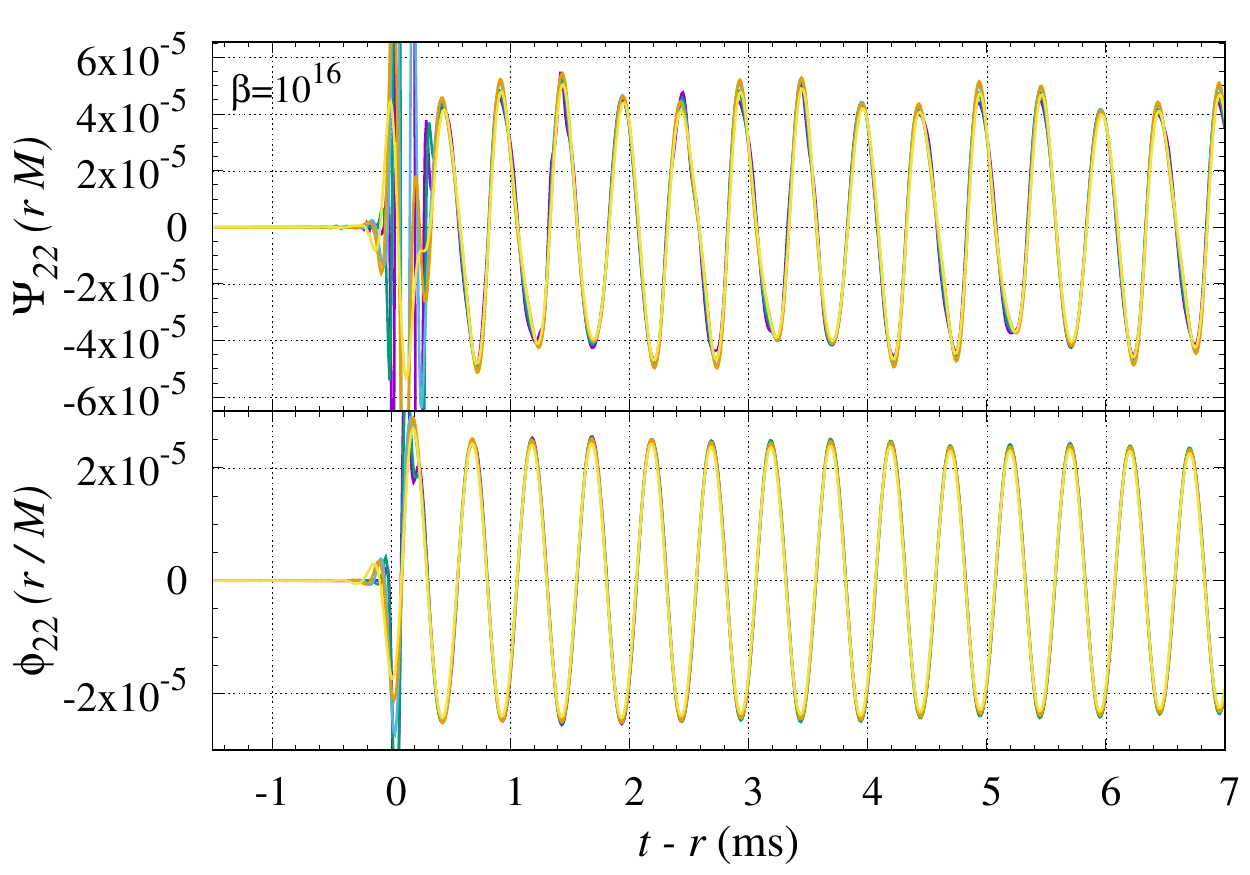}~~~
\includegraphics[width=0.46\textwidth]{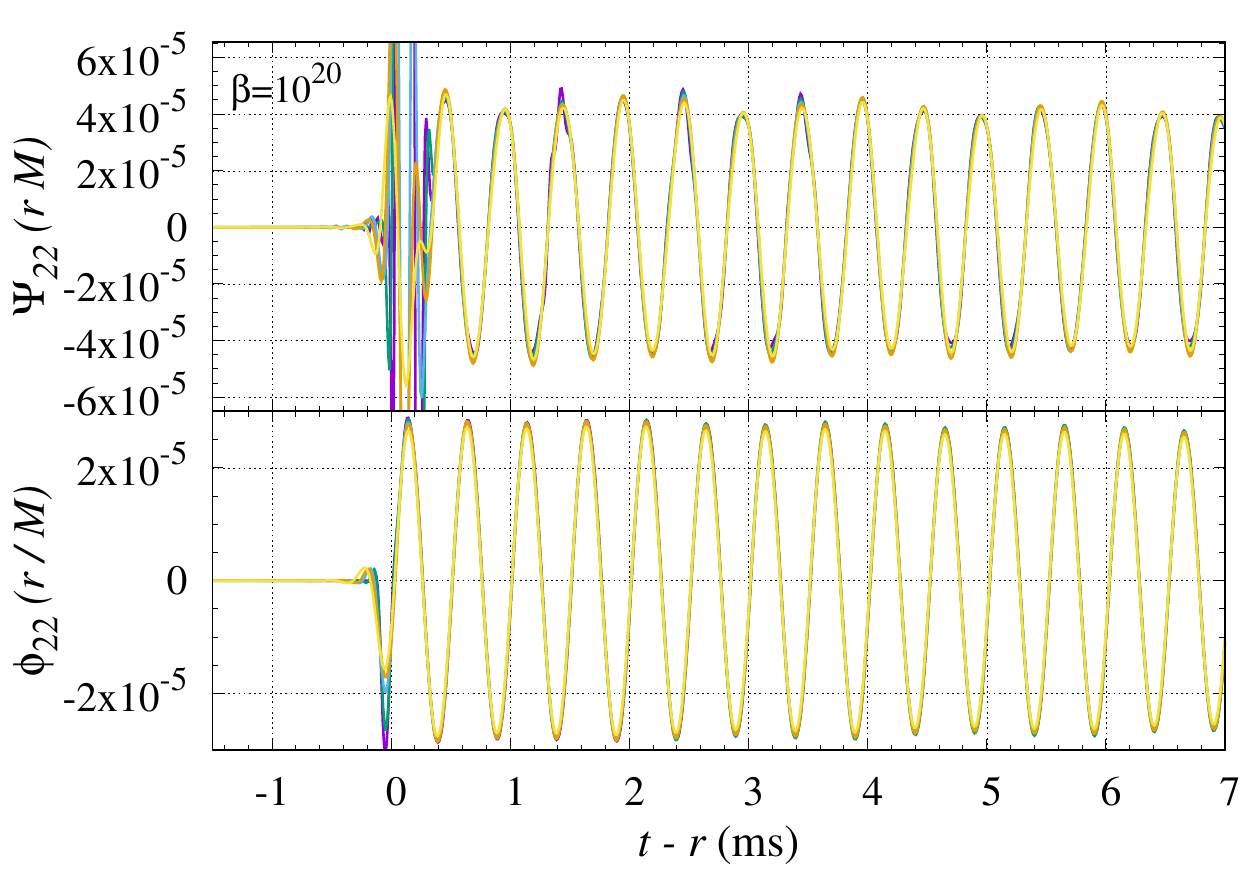}\\
\includegraphics[width=0.46\textwidth]{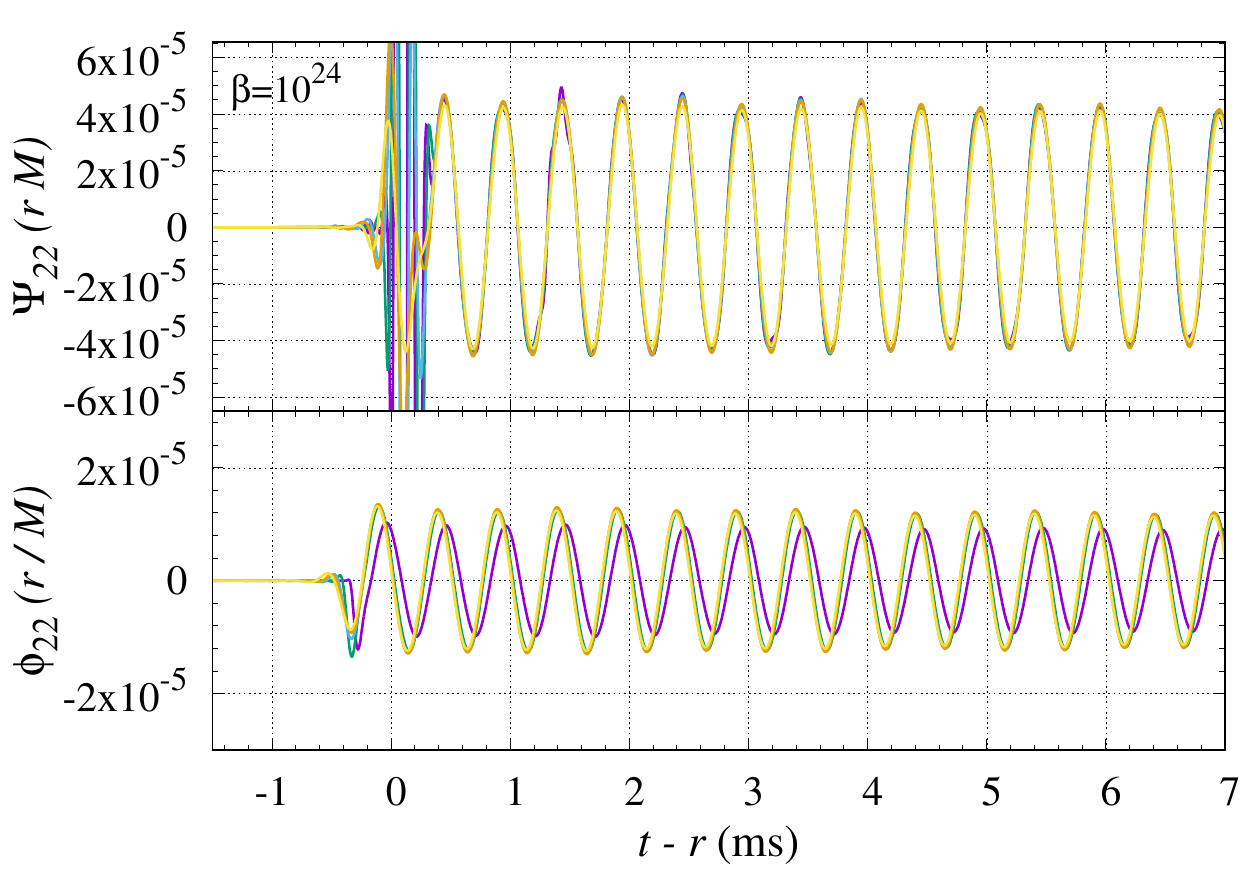}~~~
\includegraphics[width=0.46\textwidth]{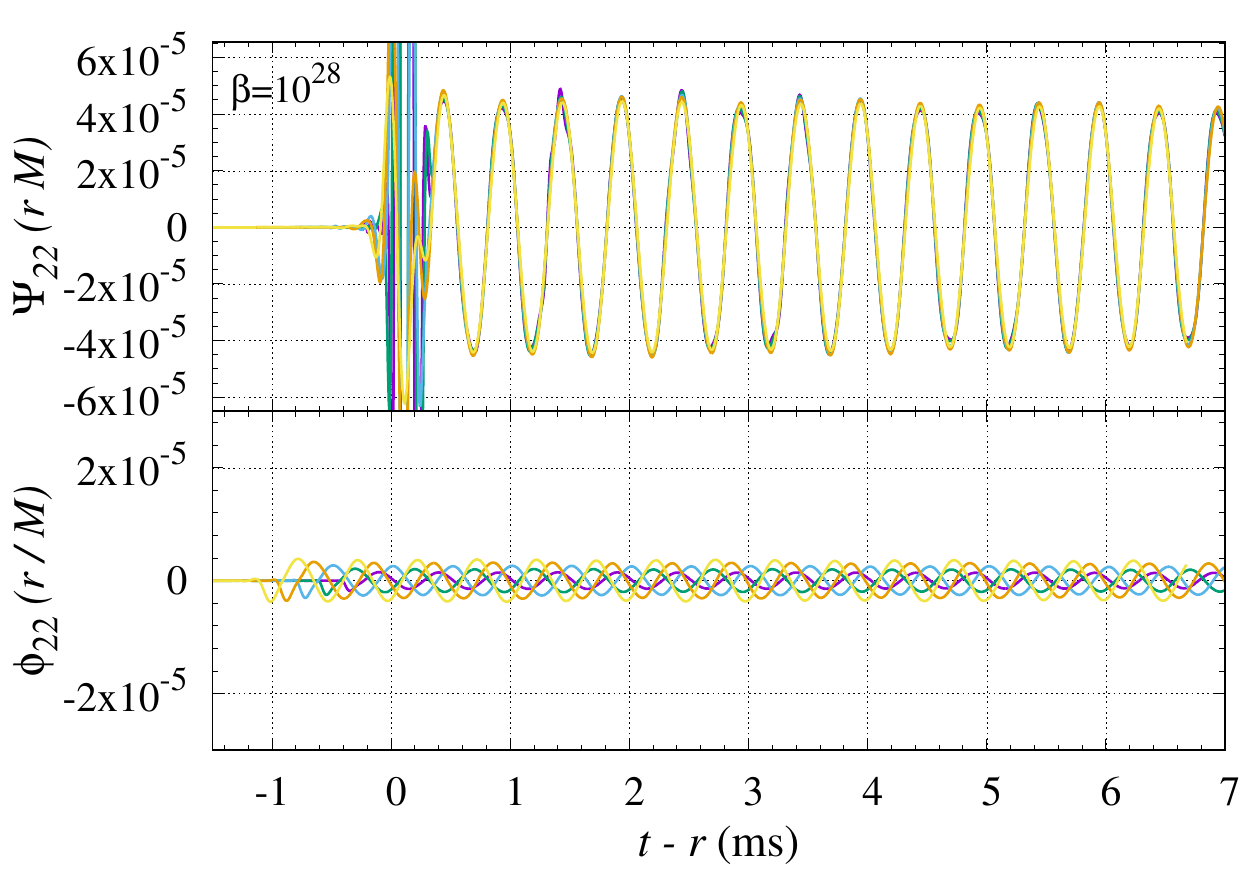}
  \caption{Gravitational and scalar waveforms of the ($l=m=2$)
    quadrupole mode as functions of the retarded time, $t-r_{\rm ex}$,
    in GR (upper left), for $\beta=1$ (upper right),
    $10^{16}$ (middle left), $10^{20}$ (middle right), $10^{24}$
    (lower left), and $10^{28}$ (lower right). For gravitational
    waves, we plot the real part of the complex Weyl scalar,
    $\Psi_{22}$. For each panel, the waveforms are plotted with
    several extraction radius, $r_{\rm ex} \approx 236$ (magenta), 354
    (green), 472 (blue), 591 (orange), and 709\,km (yellow).  (For the
    upper and middle panels as well as for gravitational waves, all
    the curves approximately overlap with each other.)  Note that the
    high amplitude waves found at $t-r \approx 0$ are the junk
    radiation numerically induced during the relaxation of the given
    initial data to those fitted to the computational setting.  Note
    that the vertical scale is the same for all the panels.
\label{fig:extraction}}
\end{figure*}


Here we analyse multipole components of scalar and gravitational waves. 
For the scalar waves, we define
\beq
\phi_{lm}={\rm Re}\left(\oint d\cos\theta d\varphi (\phi-1) Y_{lm}(\theta,\varphi)
\right)\,,
\eeq
where $Y_{lm}$ is the spherical harmonics, and pay attention to the $l=m=2$ mode. 
Gravitational waveforms are analysed by first extracting the outgoing component of the complex Weyl scalar and by decomposed into multipole modes~\cite{Yamamoto:2008js}.
Since the waves are approximately monochromatic, the gravitational wave amplitude, $h_{lm}$, may be calculated from each multipole mode of the complex Weyl scalar, $\Psi_{lm}$, by
\beq
h_{lm}=2 \omega_w^{-2}|\Psi_{lm}|\,, 
\eeq
where $\omega_w$ is the angular velocity of gravitational waves and in the present case $GM\omega_w \approx 0.087$ with $M=1.4M_\odot$.
Thus, $h_{lm}(r/M) \approx 260 |\Psi_{lm}| (r M)$. 


Figure~\ref{fig:extraction} plots the quadrupole waveforms of gravitational and scalar waves for $\beta=1$, $10^{16}$, $10^{20}$, $10^{24}$, and $10^{28}$, as well as in GR.
The amplitude of $\Psi_{22} (r_{\rm ex}M)$ is $\sim 4 \times 10^{-5}$ irrespective of the value of $\beta$.
For scalar waves, if the condition, $r_{\rm sc} \alt \lambda_{\rm wave}$, is satisfied, the asymptotic amplitude is $\phi_{22}(r_{\rm ex}/M)=(2$--$3)\times 10^{-5}$.
The order of magnitude for this agrees approximately with the expected value calculated by $(M_{\rm S}/M)(v/c)^2$ where $v \sim \sigma R \sim c/30$ with $R$ the stellar radius and $M_{\rm S}/M \sim \alpha_{\rm s}^2=10^{-2}$.
In addition, the waveforms with different extraction radii as functions of the retarded time, $t-r_{\rm ex}$, approximately align with each other for the case where $r_{\rm sc} \alt \lambda_{\rm wave}$.
This behaviour is always found for gravitational waves irrespective of the screening effect. 

By contrast, for $r_{\rm sc} \agt \lambda_{\rm wave}$, the amplitude defined by $\phi_{22} (r_{\rm ex}/M)$ increases with the extraction radius whenever $r_{\rm ex} \alt r_{\rm sc}$.
Moreover, the waveforms with different extraction radii as functions of the retarded time, $t-r_{\rm ex}$, do not overlap for this case, because of the presence of a large factor of $F(X) \gg 1$ in the screening region (see the scalar waveforms for $\beta=10^{24}$ and $10^{28}$).
To determine the asymptotic amplitude of scalar waves, we have to extract them in a far zone, in which $F(X) \approx 1$ or we perform an extrapolation.
In the present work, we consider the latter possibility for high values of $\beta \geq 10^{28}$.

\begin{figure}[t]
\includegraphics[width=0.45\textwidth]{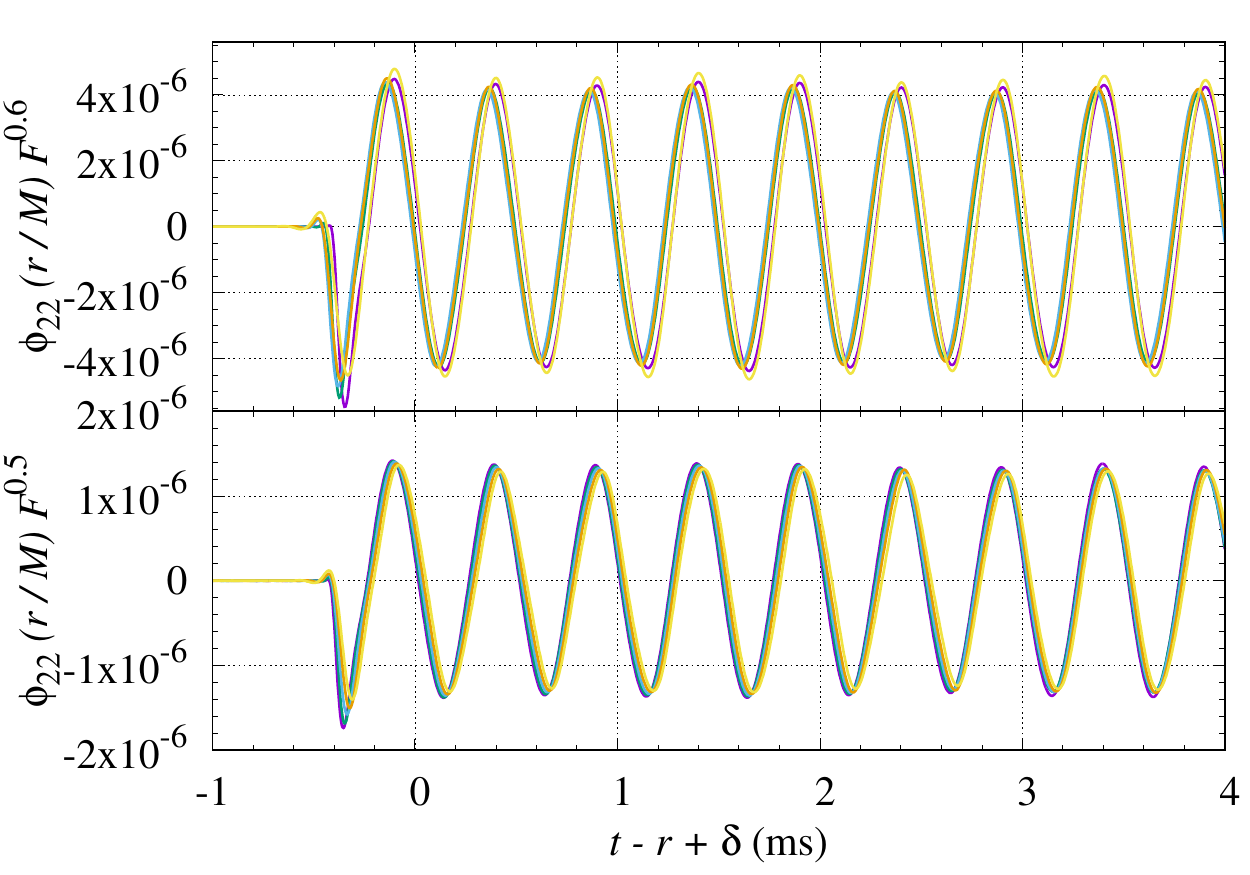}
\caption{The scalar waveform of $\phi_{22}(r_{\rm ext}/M) F^\eta$ as a function of the retarded time for $\beta=10^{28}$ with $\eta=0.6$ (upper) and $\beta=10^{32}$ with $\eta=0.5$ (lower). The extraction radius and the meaning of the colour are same as in Fig.~\ref{fig:extraction}.
To align the waveforms, the plots for $r_{\rm ex} \approx 354$ (green), 472 (blue), 591 (orange), and 709\,km (yellow) are shifted to the positive time direction.}
\label{fig:scalar_waveforms}
\end{figure}

Since $F(X)$ decreases approximately proportional to $r^{-n}$ with $n\approx 1.6$ outside the neutron stars (see Fig.~\ref{fig:spher_NS}), it is possible to predict the behaviour of the amplitude for $\lambda_{\rm wave} \alt r \alt r_{\rm sc}$ using the following method.
Neglecting the curvature effect (i.e., assuming the flat spacetime), approximating $F$ as a fixed background and setting $T=0$, the equation of $\phi_{lm}$ 
can be written as
\beqn
\left[ -\pa_t^2 +\pa_r^2 + \frac{2-n}{r} \pa_r - \frac{l(l+1)}{r^2}\right]
\phi_{lm}=0\,. \label{philm}
\eeqn
In addition, we assume that $\phi_{lm} \propto \exp(i \omega_{sw} t)$.
The general solution of Eq.~\eqref{philm} is written in terms of the outgoing component of the modified Bessel function, $Z_\nu$,
\beq
\phi_{lm}=r^{(n-1)/2} Z_\nu (\omega_{sw} r) \exp(i\omega_{sw} t)\,, \label{ZZ}
\eeq
where $\nu=\sqrt{l(l+1)+(n-1)^2/4}$.
Since the amplitude of $Z_\nu$ is proportional to $r^{-1/2}$ for $\omega_{sw} r \gg 1$ irrespective of $\nu$, the wave amplitude of $\phi_{lm}$ is proportional to $r^{n/2-1}$.
Thus for $n=1.6$, $\phi_{lm} \propto r^{-0.2}$, which implies that $\phi_{lm}(r_{\rm ex}/M) \propto r_{\rm ex}^{0.8} \propto
F^{-1/2}$; i.e., the amplitude defined by $\phi_{lm}(r_{\rm ex}/M)$ increases with radius.
For $\beta=10^{28}$, we find that $\phi_{22}(r_{\rm ex}/M)$ is approximately proportional to $F(X)^{-0.6}$ of $r=r_{\rm ex}$ because for this case, the screening region with $r > \lambda_{\rm wave}$ is rather narrow.
However, for $\beta=10^{30}$ and $\beta=10^{32}$, we confirm that the relation of $\phi_{22}(r_{\rm ex}/M) \propto F^{-1/2}$ is satisfied well.
Thus, for determining the asymptotic amplitude in these cases, we utilise this approximate relation. 
In Fig.~\ref{fig:scalar_waveforms}, we plot $\phi_{lm}(r_{\rm ex}/M)F^{\eta}$ as a function of the retarded time
with $\eta=0.6$ and 0.5 for $\beta=10^{28}$ and $10^{32}$, respectively, showing that this extraction method works well.

The analysis performed here also gives the reason that a phase misalignment is found among the scalar waves of different values of $\beta$ plotted in Fig.~\ref{fig:22pole_waveform} and among those with the different extraction radii for $\beta \geq 10^{24}$ plotted in Fig.~\ref{fig:extraction}. As described in Eq.~(\ref{ZZ}), the wave phase is determined by the functional form of $Z_\nu$. Thus during the propagation of scalar waves in a region of $F >1$, the wave phase is changed and this is reflected in the asymptotic wave phase.

We note that not only for $M_{\rm T}=1.4M_\odot$ but also for other values of $M_{\rm T}$ we find that $F \propto r^{-1.6}$ is approximately satisfied outside the neutron star with $r < r_{\rm sc}$.
Thus, the analysis shown here is likely to be valid for any neutron star. 

The above analysis also shows that for $n < 0$, the amplitude defined by $\phi_{lm}(r_{\rm ex}/M)$ decreases with the radius. Thus, if a wave is generated in $r \alt 8$\,km, the wave amplitude is suppressed, and hence, the wave amplitude should depend strongly on the wave generation region. 

\bibliography{biblio}

\end{document}